\documentclass{aa}

\usepackage{graphicx} 
\usepackage[varg]{txfonts}
\usepackage[colorlinks=true,allcolors=blue]{hyperref}%
\usepackage{lineno}

\renewcommand*\vec[1]{\ensuremath{\mathbf{#1}}}

\newcommand{\eq}[1]{Eq.~(\ref{eq:#1})}

 \newcommand{\sect}[1]{Sect.~\ref{sec:#1}}

\newcommand{\fig}[1]{Fig.~\ref{fig:#1}}
\newcommand{\Fig}[1]{Figure~\ref{fig:#1}}

\newcommand{\llnr}[1]{{\bf \color{magenta}{[]}} \color{black}} 

\defcitealias{Karpen17}{K17}
\begin{document}

\title{Propagation of untwisting solar jets from the low-beta corona into the super-Alfvénic wind: testing a solar origin scenario for switchbacks}
 
\titlerunning{Propagation of untwisting solar jets toward the super-Alfvénic solar wind}
\authorrunning{Touresse et al.}

 \author{
 J. Touresse \inst{1}\orcid{0009-0003-5268-5128}
 E. Pariat\inst{1,2}\orcid{0000-0002-2900-0608}, 
 C. Froment \inst{3}\orcid{0000-0001-5315-2890}
 V. Aslanyan\inst{4}\orcid{0000-0003-3704-4229}
 P. F. Wyper\inst{5}\orcid{0000-0002-6442-7818} \and
 L. Seyfritz \inst{1}\orcid{0009-0006-9335-6947} 
 }

 \institute{Sorbonne Universit\'e, \'Ecole polytechnique, Institut Polytechnique de Paris, Universit\'e Paris Saclay, Observatoire de Paris-PSL, CNRS, Laboratoire de Physique des Plasmas (LPP), 75005 Paris, France\\ \email{jade.touresse@lpp.polytechnique.fr}\\
        \and French-Spanish Laboratory for Astrophysics in Canarias (FSLAC), CNRS IRL 2009, Instituto de Astrofísica de Canarias, 38205, La Laguna, Tenerife, Spain\\
                 \and LPC2E, OSUC, Univ Orleans, CNRS, CNES, F-45071 Orleans, France\\
      \and School of Mathematics, University of Dundee, Dundee, DD1 4HN, UK\\
                 \and Durham University, Department of Mathematical Sciences, Stockton Road, Durham, DH1 3LE, UK}

\date{Received / Accepted }

\abstract 
{Parker Solar Probe's (PSP) discovery of the prevalence of switchbacks (SBs), localised magnetic deflections in the nascent solar wind, has sparked interest in uncovering their origins. A prominent theory suggests these SBs originate in the lower corona through magnetic reconnection processes, closely linked to solar jet phenomena. Jets are impulsive phenomena, observed at various scales in different solar atmosphere layers, associated with the release of magnetic twist and helicity.}
{This study examines whether self-consistent jets can form and propagate into the super-Alfvénic wind, assesses the impact of different Parker solar wind profiles on jet dynamics, and determines if jet-induced magnetic untwisting waves display signatures typical of SBs.}
{We employed parametric 3D numerical magnetohydrodynamics (MHD) simulations using the Adaptively Refined Magnetohydrodynamics Solver (ARMS) code to model the self-consistent generation of solar jets. Our study focuses on the propagation of solar jets in distinct atmospheric plasma $\beta$ and Alfvén velocity profiles, including a Parker solar wind. We explored the influence of different atmospheric properties thanks to analysis techniques such as radius-time diagrams and synthetic in situ velocity and magnetic field measurements, akin to those observed by PSP or Solar Orbiter.}
{Our findings demonstrate that self-consistent coronal jets can form and then propagate into the super-Alfvénic wind. Notable structures such as the leading Alfvénic wave and trailing dense-jet region were consistently observed across different plasma $\beta$ atmospheres. The jet propagation dynamics are significantly influenced by atmospheric variations, with changes in Alfvén velocity profiles affecting the group velocity and propagation ratio of the leading and trailing structures. U-loops, which are prevalent at jet onset, do not persist in the low-$\beta$ corona but magnetic untwisting waves associated with jets exhibit SB-like signatures. However, full-reversal SBs were not observed.}
{These findings may explain the absence of full reversal SBs in the sub-Alfvénic wind and illustrate the propagation of magnetic deflections through jet-like events, shedding light on possible SB formation processes.}


\maketitle

\section{Introduction} \label{sec:Introduction}

Driven by the unprecedented observations of the Parker Solar Probe \citep[PSP ; ][]{Fox16} and Solar Orbiter \citep[Solar Orbiter ;][]{Muller20,Garcia21} space missions,  scientific questions linking solar coronal phenomena and structures with in situ measurements in the inner heliosphere are receiving a tremendous amount of attention \citep{Raouafi23}. Among those outstanding questions is the possible link between coronal jet-like events with solar wind (SW) switchbacks (SBs). 

The SB phenomena have attracted significant attention from the community since the launch of PSP due to their ubiquitousness in the PSP measurements at a few tens of solar radii \citep{Raouafi23}, while they were only seldomly observed at the Earth orbit and beyond by previous missions \citep[e.g.][]{Yamauchi04, Landi06, Gosling09, Matteini14}. SBs are intermittent events characterised by a sharp deflection of the magnetic field vector away from the ambient direction and back. While most have deflections $< 90\degr$ \citep{DudokdeWit20,Bandyopadhyay22}, some do lead to full reversals, and are called full-reversal SB hereafter. We diverge from definitions given in other studies that restrict SBs to instances with Alfvénic inversion (i.e. magnetic deflections exceeding $90\degr$). This is supported by studies such as \citet{DudokdeWit20} and \citet{Bizien23}, which provide evidence that SBs are self-similar structures. This characteristic implies that the properties of SBs — such as the type of boundary or their thickness — are not significantly affected by the magnitude of the magnetic deflection. The SBs are Alfvénic features in which the deflection is most noticeable in the radial magnetic field component (but keeping a roughly constant normal magnetic field intensity $|\vec{B}|\sim$ constant). They are associated with an increase in the SW radial velocity, $v_r$. The SBs have thus also been associated with plasma jets \citep{Bale19,Kasper19}. A central open question about SBs is their formation mechanism. Two main families of scenarios have emerged: whether SBs are formed in situ in the SW \citep{Ruffolo20,Squire20,Squire22,Toth23}, or whether they find their source low in the solar atmosphere \citep{Drake21,Wyper22,Bale23}. Building on the latter scenario, the question is then which solar features and events would be the SBs'  progenitors.
Solar jet-like features appear as a putative SB source. Similarly to SB, solar jets are both intermittent and ubiquitous phenomena, prominent in the low solar atmosphere. Jets are a middle range aspect of solar activity, in terms of temporal, spatial, and energetic scales. They can be generally defined as impulsive collimated outwardly propagating transient features, observed in the solar atmosphere in emission or absorption in a wide range of spectral lines (H$\alpha$ through soft X-rays), temperatures ($10^4- 10^7$ K), and spatial scales ($10^3-10^8$ m). Depending on their size, locations, and waveband of observation, solar jet-like events have historically received a variety of names, such as spicules, photospheric jets, chromospheric jets, surges, jetlets, macrospicules, coronal jets, extreme-ultraviolet (EUV) jets, X-ray jets, among others \citep[cf. reviews of][]{Raouafi16,ShenY21}. In terms of Sun-heliosphere relationships, key questions regarding solar jets are whether some by-products of some of these jet-like events are able to reach the inner heliosphere, what would then be the in situ signature of these events, and would those signatures correspond to the SBs' properties.
It is evident that jets observed in the closed corona, such as those present in the vicinity of active regions \citep{Nishizuka11}, have little chance to reach the SW. There is, however, a tremendous amount of spicules, jetlets, and jets present in coronal holes \citep[e.g.][]{Savcheva07,Raouafi08,Raouafi14,Shimojo09,Kumar22,Skirvin22}, which are thus within the open magnetic field regions magnetically connected to the (fast) SW. Some large-scale jets have been observed to indeed propagate over a few solar radii away from the solar surface \citep{WangYM98b,Patsourakos08,Nistico09,Pucci13,Moore15}. The S-shaped structure in coronagraphic data observed by \citet{Telloni22} may also be the signature of a particularly large jet event propagating upwards. However, such a coronagraphic detection has been relatively infrequent, regarding the jet frequency. Using $70$ minute multi-site observations of solar eclipse white light images, hence with an unprecedented image quality and cadence compared to standard coronagraphic images, \citet{Hanaoka18} show that all EUV jets whose brightnesses are comparable to ordinary soft X-ray jets and that occurred in the polar regions near the eclipse period, could be seen in eclipse images propagating further than 2 $R_\sun$ (in heliocentric distance). They conclude  that ordinary polar jets shall generally reach high altitudes and escape from the Sun as part of the SW. Their results tend to indicate that our capacity to detect jet propagation in coronal images (either direct or by coronagraph) is strongly instrument limited. The absence of smaller jets (e.g. jetlets) being detected by imaging instruments likely does not mean that such events do not propagate far away towards the SW.
Based solely on solar observations, some studies speculate that SBs and microstreams can be induced by jet-like events \citep{Neugebauer95,Neugebauer12,Sterling20,HuangN23}.
Eruptions and coronal mass ejections (CME), which, being major perturbations, can be easily tracked from their triggering site in the low solar corona by EUV imagers, through the middle corona and the inner heliosphere thanks to (heliospheric) coronagraphs, and detected by in situ instruments. In contrast, jets, even the largest of them, are too small to be unambiguously associated with a specific in situ feature. To date, the study by \citet{Parenti21} is the sole study to have achieved a deterministic one-to-one association between specific coronal jet observations and in situ measurements. This underscores the inherent challenges of performing such a direct comparison, even for large jets, particularly in terms of connectivity.

Due to gaps in observations and the community-limited capacity to model, sufficiently precisely, the magnetic connection between a particular region of the heliosphere with its counterpart on the Sun, the link between jet-like events and SBs can, at best, be done statistically, and even this approach is not straightforward. Several authors have compared SBs' statistical properties during PSP perihelion, trying to link them with the statistical distribution of structures and events in the low solar atmosphere \citep{Fargette21,DePablos22,Raouafi23,Kumar23,HouC24,Bizien24}. Using wavelet transform of magnetic field measurements, \citet{Fargette21} note that SB occurrence and spectral properties appear to depend on the source region of the SW rather than on the radial distance of PSP. They also found periodic spatial modulations consistent with solar granulation and super granulation, suggesting an influence of the low solar corona on the SB formation mechanism. \citet{Bale23} show that the mixed-polarity photospheric radial magnetic field distribution is on the same spatial scale as microstreams' events observed by PSP, suggesting the importance of interchange reconnection in the formation process. Focusing on a corotation interval, \citet{Kumar23} identified recurring jets, arising from interchange or breakout reconnection at coronal hole bright points and plume bases, as the probable origins of microstreams and SBs observed in the SW.
Up to now, the state of the art of numerical modelling is similarly limited in its capacity to deterministically link low corona dynamics with SBs. 'All-inclusive' simulations of the SW formation are only nascent \citep{Iijima23} and are not yet able to resolve all the necessary structures. With two-dimensional (2D) particle-in-cell (PIC) simulations, \citet{Bale23} point out the potentially important role of interchange reconnection dynamics in explaining SBs. Using three-dimensional (3D) magnetohydrodynamic (MHD) simulations, \citet{Wyper22} emulated flythroughs within an interchange reconnection-generated curtain of propagating and interacting torsional Alfvénic waves and found Alfvénic patches that closely resemble observations of SBs, although with relatively small magnetic field deflections. In these models, it is directly the properties of interchange reconnection that dictate the properties of the SBs. However, the modelled current sheets are unrealistically large, and how the properties of the Alfvénic patch vary with more realistic coronal current sheets remains to be explored.

Focusing on jet-like events as precursors for SBs, only a few simulations have been produced and analysed \citep{Lionello16,Karpen17,Szente17}. This is surprising given the fact that jet-like event modelling is a mature field that has benefited from numerous studies \citep[cf. reviews][]{Raouafi16,ShenY21,Skirvin22,Tziotziou23}. Jet modelling has mostly been focused on understanding the trigger process of jets on the complex dynamics of the multi-thermal and multi-flows' dynamics, in comparison to coronal observational features, and particle accelerations \citep[see e.g.][for recent models]{Pariat16,Wyper19,GonzalezAviles21,Pallister21,ChenF22,ZhuJ23}. Jet modelling has demonstrated that the observed features are induced by several simultaneously acting physical processes such as reconnection-driven outflows \citep[exhaust from the reconnection site; e.g.][]{Yokoyama95,MorenoInsertis08,Archontis10a}, evaporation flows induced by heating \citep[][]{Yokoyama96,Shimojo01,Miyagoshi03,Miyagoshi04}, and the non-linear Alfvénic magnetic untwisting waves induced by reconnection between twisted field lines with untwisted ones \citep{Pariat09a,Archontis13a,Lee15,Wyper18b,Doyle19}. Three-dimensional MHD models tend to indicate that the magnetic untwisting is the energetically dominant process \citep{Pariat09a,Pariat16,FangF14}. To a large extent, jets can be seen as eruptive processes induced by reconnection-driven full destruction of a twisted structure or flux rope \citep{Pariat15a,Wyper16a,Wyper17}.

The importance of the magnetic untwisting mechanism is further enhanced in the jet propagation simulations of \citet{Lionello16}, \citet[][ hereafter \citetalias{Karpen17}]{Karpen17}, and \citet{Szente17}, despite their marked differences. The jet propagation simulations of \citet{Lionello16} modelled the corona using a spherically symmetric, steady-state solar wind resulting from the relaxation of an initial thermodynamic solution \citep[based on][]{Lionello09}. A series of jets is generated thanks to interchange between an emerging flux twisted flux rope and the ambient open coronal field. Both in \citet{Szente17} and \citetalias{Karpen17}, the jet was impulsively and self-consistently generated following photospheric-boundary shearing motions \citep[based on the model of][]{Pariat09a}. While the simulation of \citet{Szente17} relies on the  two-temperature (protons and electrons) Alfvén Wave Solar atmosphere Model (AWSoM) framework \citep{vanderHolst14}, in which the atmosphere and SW  is self-consistent induced by low-Alfvén wave turbulence, in \citetalias{Karpen17} the atmosphere is more directly initiated with a uni-thermal Parker solar wind solution. Local coronal simulations by \citet{Lionello16}, \citet{Pariat16}, \citet{Uritsky17}, \citetalias{Karpen17}, and \citet{Szente17} have identified that the dominant outcome of jet initiation and the fastest propagating structure is the Alfvénic untwisting magnetic wave. Moreover, \citet{Pariat16} and \citet{Uritsky17} found that the propagation of these waves leads to significant complexity, characterised by various structures such as the wave front, a shear Alfvén turbulence domain, shear and compressible turbulence, and ultimately a dense jet. \citet{Roberts18} carried out an analysis of the possible signatures of the \citetalias{Karpen17}'s jet as it was upwardly propagating, and produced synthetic signatures that would be observed by in situ instruments measuring the magnetic and velocity fields. The result being that the untwisting wave front does indeed correspond to a local magnetic deflection, with a decrease in the radial magnetic field intensity, $|B_r|$, while the norm of the magnetic field $|B|$ remained roughly constant, and that the untwisting wave was also associated with enhanced radial velocity \citep[cf. Fig. 9 of][]{Roberts18}.

Although only a single jet was analysed, the study by \citet{Roberts18} suggests that some SBs and microstreams can indeed result from magnetic untwisting waves generated along solar jet-like events. However, the simulation of \citetalias{Karpen17} analysed by \citet{Roberts18} corresponds to a very specific atmospheric condition in which the atmosphere is magnetically dominated ($\beta <<1$) throughout the domain. At the top of the domain, at $9 R_\sun$, the value of $\beta$ in the SW is still lower than $10^{-2}$. The wind is sub-Alfvénic everywhere in the domain. The Alfvén surface, defined as the radially increasing SW speed, equals the local Alfvén speed \citep{Cranmer23}, and through which the solar wind transition from sub- to super-Alfvénic is thus absent from the setup of \citetalias{Karpen17}. Furthermore, the simulation case of \citetalias{Karpen17} corresponds to a rather extreme case of an open corona, in which the Alfvén surface would be tens of solar radii away (at $~45-50 R_\sun$), which is larger than the most recent models and observations of the range of location of the Alfvén surface \citep[cf. Sect 3 of the review][and reference therein]{Cranmer23}.

The objective of the present study is thus to build upon the works of \citet{Pariat16}, \citetalias{Karpen17}, and \citet{Roberts18} to analyse the propagation of coronal jets into the solar wind, by performing parametric simulations using coronal profile that includes an Alfvénic surface inside the simulation domain. Our objective is to determine whether and how the outputs of the self-consistent generation of a coronal jet are influenced by the plasma $\beta$ properties of the background atmosphere. Such an analysis was partly carried by \citet{Pariat16}, but restricted to a uniform atmosphere (without solar wind and gravity) and in a domain limited to a few fractions of the solar radius above the surface. \citet{Pariat16} find that the background plasma $\beta$ significantly influences the outward propagation of the wave, with a clear separation between the Alfvénic magnetic untwisting wave and the dense jet at low $\beta$, whereas the two were embedded within each other when the background was uniform and with $\beta \sim 1$. When studying the propagation over several solar radii, the question now is how the propagation of the jet evolves with a radially changing background plasma $\beta$.

In the present paper, we want to address the following questions: 
\begin{itemize}
    \item The generation and propagation of self-consistent jets towards the super-Alfvénic wind,
    \item The influence of different Parker solar wind profiles on jet propagation,
    \item The investigation into whether the magnetic untwisting wave, triggered by a jet-like event, results in signatures typically associated with SBs or full reversal SBs.
\end{itemize}
 Our paper is organised as follows. In \sect{Sim} we first summarise the concept and properties of the numerical experiments. In \sect{jet_propagation} we study the dynamics of different propagating structures revealed through parametric simulations. In \sect{mag_deflections} we focus on one specific structure of the jet, a leading Alfvénic wave, and examine whether its characteristics align with those of a SB. Finally, in \sect{Conclusion} we summarise our results and discuss them in the broader context of the relation between jets and a SB.

\section{Simulation description} \label{sec:Sim}

This section begins with an introduction to the numerical domain (\sect{Num}), laying the groundwork for our consecutive analyses. Subsequently, we will outline the initial conditions (\sect{Init}), setting the stage for three distinct parametric simulations (\sect{param_sim}). Finally, we will present the first main phases pivotal to the generation of the jet, offering a comprehensive overview of our scientific study (\sect{Steps}). 

\subsection{Numerical model} \label{sec:Num}

The three-dimensional numerical simulations are performed using the Adaptively Refined Magnetohydrodynamic Solver \citep[ARMS,][]{DeVore91,DeVore08} to solve time-dependent, isothermal, compressible, ideal MHD equations within a spherical coordinate framework, akin to the approach presented in \citetalias{Karpen17} (cf. Sect. 2 and the model therein). 

Moreover, the gas follows a ideal gas law for fully ionised hydrogen: 
\begin{equation}
    P = 2 \frac{k_{B}}{m_{\mathrm{p}}}\rho T \label{eq:idealgas}
\end{equation}
with thermal pressure $P$, mass density $\rho$, proton mass $m_{\mathrm{p}}$, and the Boltzmann constant $k_{B}$. ARMS also evolves the magnetic field, $\mathbf{B}$, and the velocity field, $\mathbf{v}$. We impose a constant and uniform temperature of the plasma, $T$, and do not solve any temperature equation. 

\begin{figure*}
\sidecaption
 \includegraphics[width=12cm]{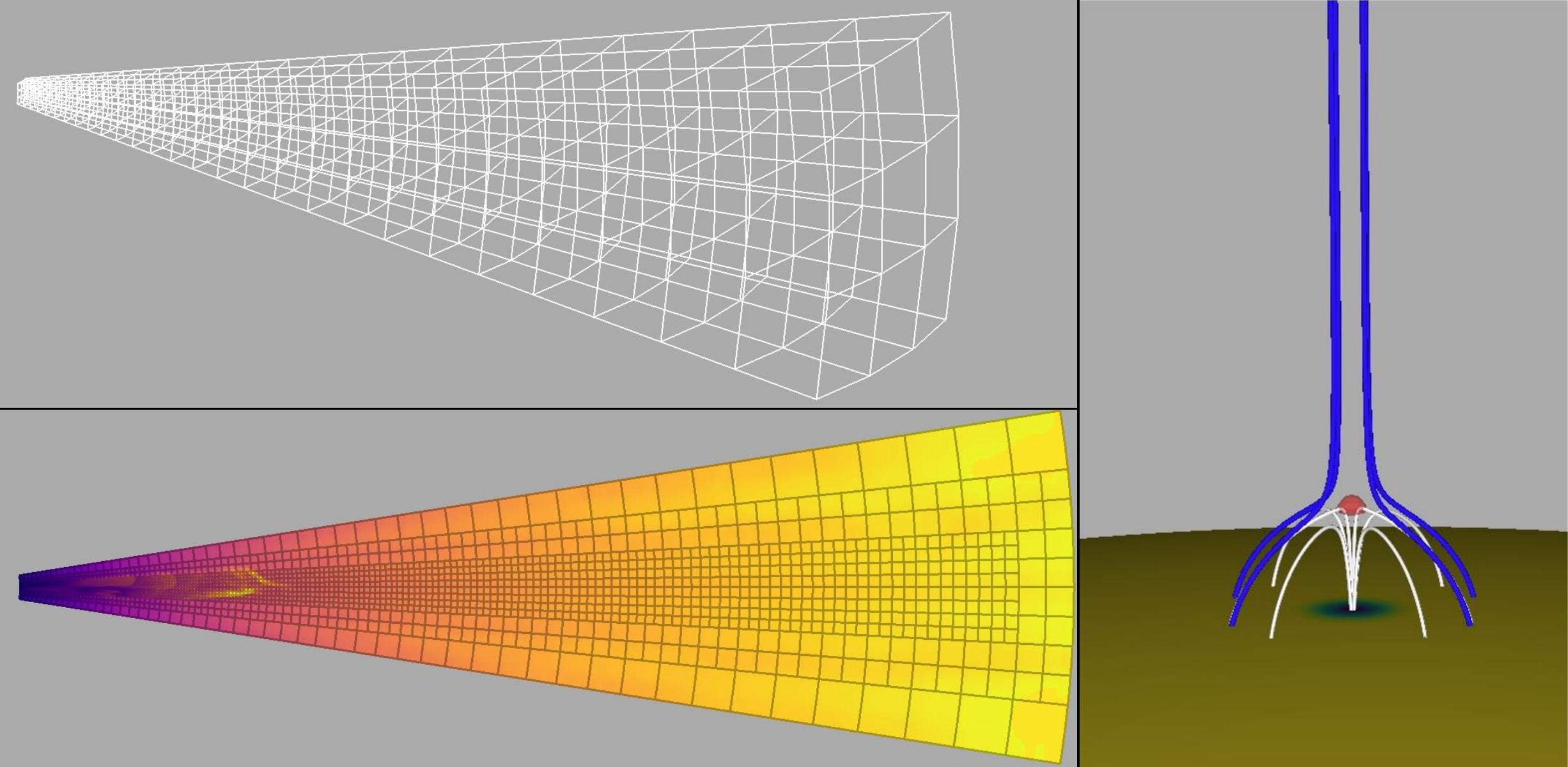}
  \caption{Simulation domain and magnetic topology. Top left: 3D volume of the domain of the simulation with the root blocks of the grid. Bottom left: A 2D cross section at constant angle of $\phi = 0 \degr$ of the velocity colour at t = $5\,500$ s. This snapshot highlights the spatial distribution of the velocity. The grid block boundaries are delineated by grey lines. Each block contains $8 \times 8 \times 8$ cells. Right: Initial magnetic topological structure. The isosurface of plasma $\beta$ = 20 (red spheroid) indicates the location of the 3D magnetic null point. Magnetic field lines of distinct connectivity bounding the separatrix surface are represented, either closed (white field lines) or open to the heliosphere (blue field lines).}
  \label{fig:domain_topo}
\end{figure*}

The ARMS code employs the flux-corrected transport (FCT) algorithm derived from \citet{DeVore91}. The FCT algorithm ensures that the magnetic field remains divergence-free within the precision limits of the machine. This approach also prevents the emergence of non-physical outcomes, such as negative mass densities. Additionally, it minimises numerical oscillations associated with intense gradients that occur at the grid scale. This code uses the PARallel Adaptive MESH refinement (PARAMESH) toolkit \citep{MacNeice00} that offers parallel adaptive mesh refinement, dynamically adjusting the grid throughout the computation to accommodate the evolving solution.

The domain of the simulation is a spherical wedge that spans radially from $1\,R_{\sun}$ (the solar surface) to $15\,R_{\sun}$, and $\pm 9 \degr$ in both co-latitude, $\theta$, and longitude, $\phi$, from a central spine axis of the domain of coordinates ($\theta = 90 \degr$, $\phi = 0 \degr$; see \fig{domain_topo}). On this domain, there is an adaptive mesh with a block-structured and dynamic grid.
The computational domain exhibits a radial-to-perpendicular ratio of $45:1$, with the outer boundary extending significantly beyond the sonic point denoted as $r_\mathrm{s}$, defined in \sect{sw}. The grid is uniformly spaced in angles and exponentially in radius. This arrangement ensures a consistent $\partial r/r$ ratio throughout various levels of refinement. The domain is built upon a framework of $ 27 \times 3 \times 3 $ grid root blocks, each containing $ 8 \times 8 \times 8$ cells. Up to four extra levels of grid refinement were allowed during the simulation, doubling the spacial resolution at each level (see \fig{domain_topo}). 

A volume, extending from the solar surface (entire coronal base) to a height of $1.77\times10^9$ cm ($2.5\times 10^{-2}\, R_{\sun}$ ), is imposed to refine to the level 4, having a grid resolution of $192 \times 192$ pixels, corresponding to a spatial resolution about $1.1 \times 10^8$ cm ($1.6 \times 10^{-3} \, R_{\sun}$) on the solar surface. The same maximum refinement level was imposed on a central region in the volume where the jet forms and propagates, a domain which extends up to $9.8 \times 10^{11}$ cm ($14 \, R_{\sun}$) in height and over [$-3\degr$, $+3\degr$] in angle with a spatial resolution going from $1.1 \times 10^8$ cm ($1.6 \times 10^{-3} \, R_{\sun}$) on the solar surface to $1.6 \times 10^8$ cm ($2.3 \times 10^{-3} \, R_{\sun}$) at $14 \, R_{\sun}$. The outermost perimeter of root blocks (within $3 \degr$ of the side boundaries) was restricted to only one additional level of refinement (level 2) beyond a height of $3.61\times 10^9$ cm ($5.2\times 10^{-2} R_{\sun}$) above the surface, to avoid overly resolving regions that are near the side boundaries and far from the central, jet-containing portion of the domain.  In the transitional zone between these two regions, the blocks undergo refinement up to level 3, ensuring a more gradual transition in refinement levels across the distinct areas (see \fig{domain_topo}). 

At the radial inner boundary, a line-tying condition is imposed. The tangential velocity, $v_\mathrm{b}$ is specified throughout the coronal base and the radial magnetic field, $B_r$, is specified throughout the domain. In the guard cells beneath the open radial inner boundary, both mass density and pressure have fixed values. At the radial outer boundary, we extrapolate the mass density using the ratio of outer to inner values in the Parker steady wind. All three velocity components are extrapolated, assuming zero-gradient conditions (free flow-through and free slip). The side boundaries (in $\theta$ and in $\phi$) are closed, reflecting with respect to the normal flow velocities ($v_{\theta}$ and $v_{\phi}$, respectively) and free slip with respect to the tangential flow velocities ($v_{\phi}$, $v_{\theta}$, and in both cases, $v_r$). Beyond all six boundaries, the mass density and the magnetic field vector are extrapolated using zero-gradient conditions.

\subsection{Initial conditions} \label{sec:Init}

This section provides an in-depth analysis of the initial conditions that lay the groundwork for our study. We begin by investigating the hydrodynamic conditions, utilising a solar wind model tailored for a stratified atmosphere (\sect{sw}). Subsequently, we turn our attention to the initial magnetic configuration (\sect{ini_mag}). The intricate interplay between these initial conditions yields three distinct parametric simulations, contributing to a comprehensive understanding of the complex dynamics at play (\sect{param_sim}).

\subsubsection{Solar wind model for a stratified atmosphere} \label{sec:sw}

Building on \citet{Masson13} and \citetalias{Karpen17}, we initialised the atmosphere using Parker's steady, spherically symmetric, supersonic wind model from \citet{Parker58}: 

\begin{equation}
    \frac{v_r^2}{c_{\mathrm{s}}^2} exp \left( 1- \frac{v_r^2}{c_{\mathrm{s}}^2} \right) = \frac{r_{\mathrm{s}}^4}{r^4} exp \left(4-4 \frac{r_{\mathrm{s}}}{r} \right)
    \label{eq:solar_wind_radial_velocity}
\end{equation}
where $v_r$ is the radial velocity, $c_{\mathrm{s}} = \sqrt{2 k_B T_{\mathrm{b}} /m_{\mathrm{p}}}$ is the isothermal sound speed, and $r_{\mathrm{s}} = \mathrm{G} M_{\sun} m_{\mathrm{p}}/4 k_B T_{\mathrm{b}}$ is the radius of the sonic point. We assumed a constant temperature for the three parametric simulations : $T_{\mathrm{b}} = 10^6$ K (respectively $2  \times 10^6$ K for medium and high $\beta$ simulations) giving a sound speed $c_{\mathrm{s}} = 128 \ \mathrm{km}\,\mathrm{s}^{-1}$ (respectively $182 \ \mathrm{km}\, \mathrm{s}^{-1}$ for medium and high $\beta$ simulations) and $r_{\mathrm{s}}$ = $5.7 \, R_{\sun}$ (respectively $2.9 \, R_{\sun}$ for medium and high $\beta$ simulations). This isothermal approximation simplifies the representation of the solar wind but disregards the heating effects caused by reconnection-driven outflows of the jet. As a result, we cannot make precise predictions about the observable phenomena that depend on the detailed thermodynamic behaviour of the coronal plasma. To fully determine the atmosphere, we set the  pressure at the bottom boundary, $P_{\text{b}}$ (cf. values in Table \ref{tab:tab1}), which fixes the photospheric mass density thanks to Eq. \ref{eq:idealgas}. The density in the rest of the domain is determined by maintaining a constant mass flux, $\rho_{\sun} v_{r, \sun} r_{\sun}^2 = \mathrm{cst}$. This provides an isothermal atmosphere stratified in density with a radial plasma flow.

\subsubsection{Initial magnetic configuration} \label{sec:ini_mag}

Following \citet{Pariat09a} and \citetalias{Karpen17}, to self-consistently model the onset and generation of a jet occurring in small opposite sign magnetic polarity within a predominantly uniform coronal hole, we initially combine two potential magnetic fields. To model the uniform background magnetic field of a coronal hole, we adopted a Sun-centred monopole configuration given by the equation
\begin{equation}
    \boldsymbol{B_{\mathrm{m}}} = - \lvert B_{\mathrm{m}} \rvert \frac{R_{\sun}^2}{r^2} \boldsymbol{\hat{r}}
    \label{eq:background_magnetic_field}
\end{equation}
with $\vec{\hat{r}}$, the radial unit vector.
The value of the background magnetic field for each simulation can be found in Table \ref{tab:tab1}. While magnetic monopoles are theoretical entities that do not exist in reality, a practical workaround is employed to simulate the open field characteristics of a coronal hole region within a confined computational domain. This involves utilising a monopole to generate a unipolar radially decreasing magnetic field within the simulation volume.

To fit a magnetic concentration within the coronal hole like a parasitic polarity that typically serve as the origin of jets, we used a point dipole located at a depth $d = 1 \times 10^7$ cm beneath the surface, outside the computational domain, and aligned in the radial direction. These elements result in an anemone-like structure (see \fig{domain_topo}, right panel) characterised by a dome-shaped separatrix surface, hosting a 3D null point accompanied by its corresponding fan surface and two spine lines \citep{Longcope05, Pariat09a}. This magnetic configuration, featuring an embedded-dipole arrangement, aligns closely with observations of jet-like events \citep[see][]{Nistico09, Schmieder13, Liu16, Li19, Joshi20}.
The maximum radial field on the surface is $B_{\text{d}}$ = +35 G. This dipole represents a compact embedded polarity placed at the equator (colatitude, $\theta = 90 \degr$) and central meridian (longitude, $\phi = 0 \degr$) in our spherical coordinate system to ensure optimal grid resolution. 

\subsubsection{Parametric simulations}\label{sec:param_sim}
To explore the multifaceted outcomes of this interplay, we systematically varied specific parameters, leading to the derivation of three distinct parametric simulations. The varying parameters and their corresponding values are summarised in the table below:

\begin{table}[ht]
\caption{Simulation-specific initial conditions' parameters.}
\label{tab:tab1}
\centering
\begin{tabular}{c c c c c}
\hline \hline
Simulation name & $T_{\text{b}}$ {[}K{]} & $P_{\text{b}}$ {[}bar{]}   & $B_{\text{m}}$ {[}G{]} & $B_{\text{d}}$ {[}G{]} \\ \hline
Low $\beta$     & $1\times10^6$            & $2.75 \times 10^{-8}$ & $-2.5$          & $35$      \\
Medium $\beta$  & $2\times10^6$          & $3.30 \times 10^{-8}$  & $-2.0$            & $35$            \\ 
High $\beta$    & $2\times10^6$          & $5.00 \times 10^{-8}$  & $-1.5$          & $35$            \\ \hline
\end{tabular}
\tablefoot{The first two columns display hydrodynamic parameters, such as $T_{\text{b}}$, the temperature, and $P_{\text{b}}$, the pressure at the base of the atmosphere. The next two columns display magnetic field parameters: $B_{\text{m}}$, the parameter defining the background monopole magnetic field and $B_{\text{d}}$, the parameter defining the radial magnetic field intensity of the sub-surface dipole under the surface.}
\end{table}

\begin{figure*}
    \centering
   \includegraphics[width = 0.85\linewidth]{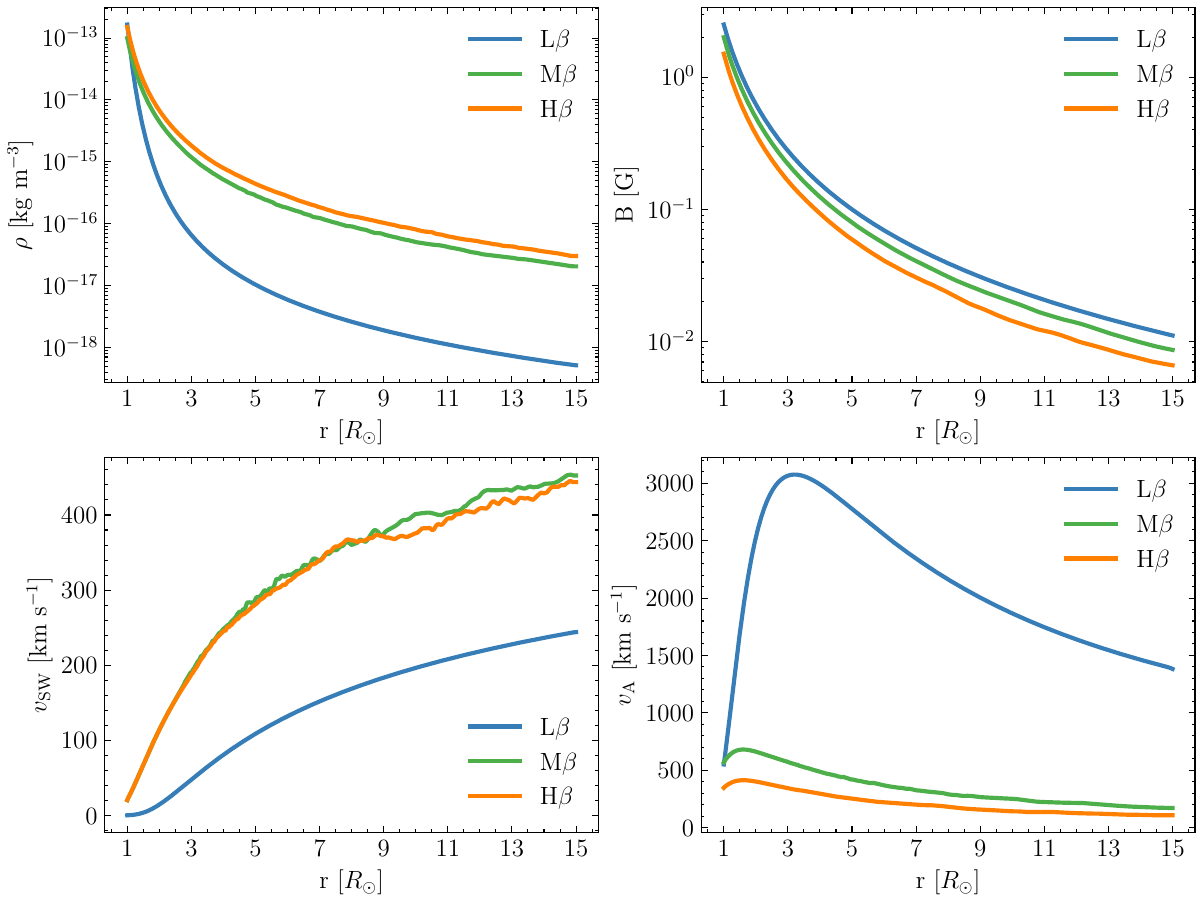} 
    \caption{Comparison of the background radial profiles of diverse quantities in the three parametric simulations: low  $\beta$ (blue curves), medium $\beta$ (green curves), and high $\beta$ (orange curves). The figure panels correspond to: mass density, $\rho$ (top left), magnetic field intensity, $|\vec{B}|$ (top right), solar wind velocity intensity, $|\vec{v_{\text{SW}}}|$ (bottom left), and Alfvén speed, $v_{\text{A}}$ (bottom right).}
    \label{fig:atmosphere_conditions_v1}
\end{figure*}

The varying choices of these parameters result in three distinctly different atmospheric models. One of these simulations, using parameters from the \citetalias{Karpen17} paper, was designated as the reference model, that is, the low $\beta$ simulation. In order to characterise these new atmospheres, we looked at the spatial evolution of different quantities as a function of the solar radius. This visual analysis is integral to our study as it enables a direct comparison of the simulated atmospheres, highlighting similarities and biases that may offer insights into common jet behaviours. To achieve optimal visualisation of the atmospheric evolution, we extract a specific line corresponding to $\theta = 97\degr$ and $\phi = 7\degr$. Selecting the central part would introduce disturbances caused by the central polarity, which could hinder the clarity of the atmospheric profile. After the relaxation was achieved, we selected key atmospheric parameters, including mass density, magnetic field, plasma $\beta$, solar wind velocity and Alfvén velocity. The subsequent paragraphs will delve into a detailed interpretation of \fig{atmosphere_conditions_v1}, elucidating the observed trends and their implications for our understanding of atmospheric phenomena.

Simulations reveal a consistent mass density pattern, displayed in the top left corner of \fig{atmosphere_conditions_v1}. Initially, there is a rapid decline within the first solar radii, followed by a slower decrease, indicating an exponential decay trend on the logarithmic scale, thus it is a double exponential curve. Medium $\beta$ and high $\beta$ curves exhibit a slower rate of decrease compared to the low $\beta$ curve, with a consistent gap between the medium and high $\beta$ curves. Despite different initial mass densities, all simulations converge to the same profile, consistent with theoretical expectations. The observed behaviours align with established models, such as $\rho v_r r^2 = \mathrm{cst}$ and $v_r \propto ln(r)^{1/2}$, hence giving the dependence $\rho \propto 1/r^2\sqrt{ln(r)}$. The similarities between the medium and high $\beta$ curves are attributed to a common temperature, while the differences between the low $\beta$ curve and the other two may result from temperature variations.

Similarly, the simulations depict a consistent magnetic field pattern, illustrated in the top right corner of \fig{atmosphere_conditions_v1}. At the solar surface, each curve aligns with the initial value of the background environment. Despite the logarithmic scale, the $1/r^2$ decay trend is evident across all curves, consistent with \eq{background_magnetic_field}.

Across all three simulations, the solar wind velocity curves exhibit a similar behaviour characterised by a logarithmic dependence, approximately resembling $ln(r)^{1/2}$. This trend aligns with the approximation of the radial velocity $v_r$ using \eq{solar_wind_radial_velocity}. However, near the solar surface, the behaviour of the low $\beta$ curve differs. The velocity seems to follow an exponential law: initially, the curve rises slowly, but then its slope increases.

Regarding the Alfvén velocity, the simulations demonstrate a consistent pattern: within the first few solar radii, there is an increase, followed by a subsequent decrease. These variations can be elucidated by the formula for the Alfvén velocity: $v_{\mathrm{A}} = B/\sqrt{\mu_0\rho}$, where $B \propto 1/r^2$ and $\rho \propto 1/r^2\sqrt{ln(r)}$, leading to $v_{\mathrm{A}} \propto {ln(r)}^{1/4}/r$. Near $r = 1 R_{\sun}$, $v_{\mathrm{A}}$ adheres to the ${ln(r)}^{1/4}$ curve, swiftly transitioning to a $1/r$ dependence. The disparity between the curves primarily stems from variations in mass density: notably, in the low $\beta$ simulation, the values at higher solar radii are significantly lower than those of the medium and high $\beta$ simulations.

\begin{figure}
     \centering
\resizebox{0.9\hsize}{!}{\includegraphics{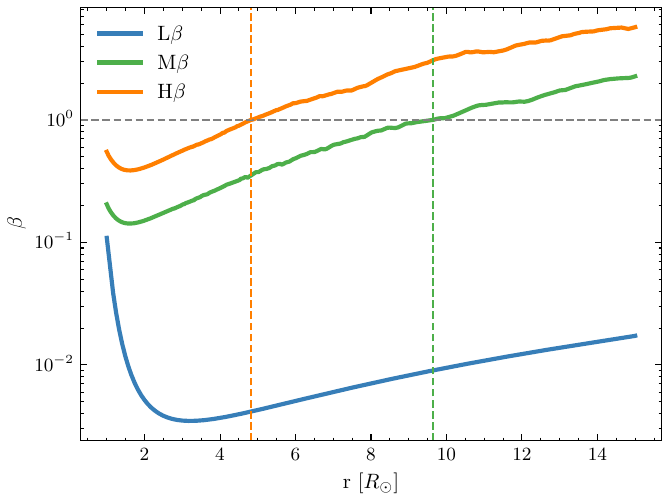} }
    \caption{Radial distribution of plasma $\beta$ for the low $\beta$ (blue), medium $\beta$ (green), and high $\beta$ (orange) simulations.}
    \label{fig:atmosphere_conditions_v2}
\end{figure}

\Fig{atmosphere_conditions_v2} displays the radial evolution of the plasma $\beta$ and explains the rationale behind the names given to each simulation. Across all three simulations, the plasma $\beta = 2 \rho k_B T \mu_0/B^2$ parameter exhibits similar variations: an initial decrease followed by an increase. These variations primarily stem from disparities in mass density values. Notably, the medium and high $\beta$ simulations display parallel trends in both mass density and plasma $\beta$ curves. The radius at which the plasma $\beta$ parameter exceeds unity is particularly significant. It marks the boundary between the region where magnetic pressure dominates (plasma $\beta \ll 1$) and the region where gas pressure dominates over magnetic pressure (plasma $\beta$ > 1).  In the so-called low $\beta$ simulation (hereafter referred to as L$\beta$), this curve does not intersect within the domain, indicating that the dynamics are dominated by magnetic pressure and tension. Conversely, in the medium and high $\beta$ simulations (hereafter referred to as M$\beta$ and H$\beta$), the curves intersect the plasma $\beta = 1$ threshold at radii of $9.6\ R_{\sun}$ and $4.8\ R_{\sun}$ respectively. Beyond these radii, the dynamics are primarily governed by environmental plasma dynamics rather than magnetic field effects.

\begin{figure}
  \centering  
\resizebox{0.785\hsize}{!}{\includegraphics{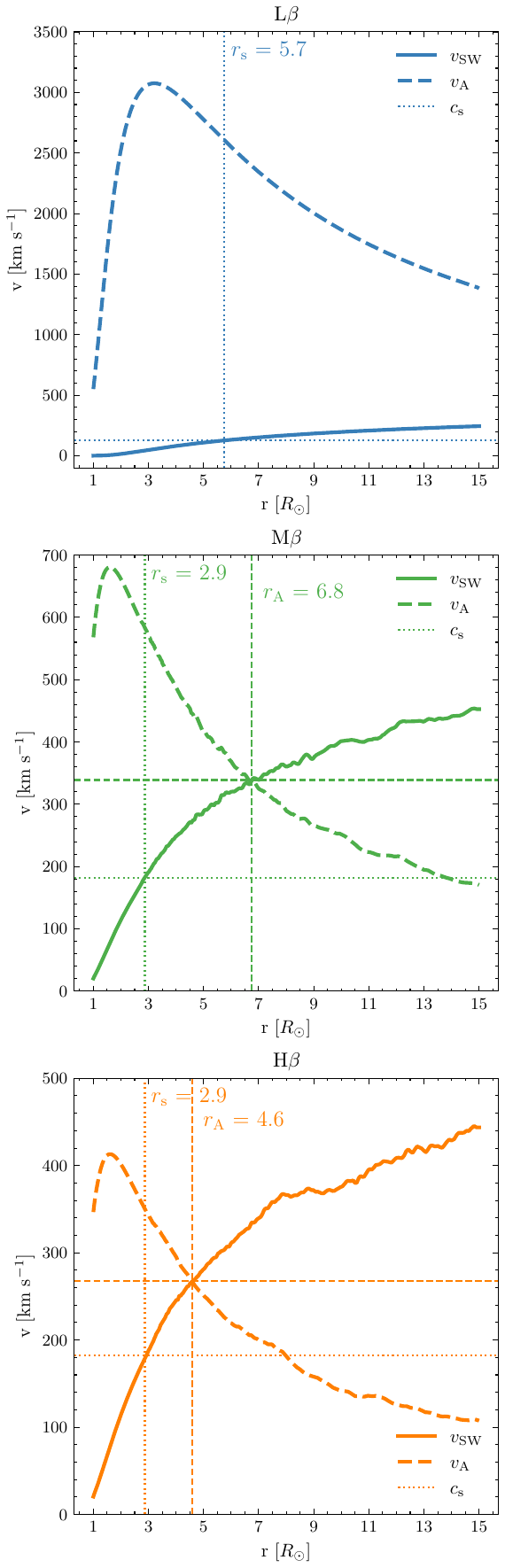}}
  \caption{Radial distributions of the atmospheric velocity, Alfvén speed, and sonic speed for L$\beta$ (top panel), M$\beta$ (middle panel), and H$\beta$ (bottom panel) simulations. The radius of the Alfvén and sonic surfaces are respectively indicated by dashed and dotted vertical lines.}
  \label{fig:comparison_velocities}
\end{figure}

\Fig{comparison_velocities} displays, for each simulation, the radial distribution of the velocity relative to the background Alfvén and sound speeds. Notably, in L$\beta$, there is no intersection between the local Alfvén velocity and the local solar wind speed, indicating a consistent sub-Alfvénic wind regime. Conversely, in M$\beta$ and H$\beta$, there exists a radius (effectively a surface) where these velocities converge. Below this threshold, the simulation remains sub-Alfvénic, while above it, the solar wind transitions to a super-Alfvénic regime. These Alfvén surfaces occur at $6.76$ and $4.59\ R_{\sun}$ respectively for M$\beta$ and H$\beta$, values consistent with the plasma $\beta$ thresholds, albeit in proportion. 

The study explores the interplay between hydrodynamic and magnetic initial conditions, resulting in three distinct atmospheric simulations. Visual analysis reveals consistent patterns in mass density, magnetic field, and solar wind velocity across simulations, while variations in Alfvén velocity and plasma $\beta$ highlight the influence of different parameters. These findings underscore the diversity of atmospheric behaviours and offer insights into the dominant factors shaping atmospheric dynamics.

\subsection{Simulation steps} \label{sec:Steps}

In this section we briefly describe the three main phases of the simulation: the initial relaxation phase (\sect{relax}), the bottom boundary driving phase (\sect{photo}) and the jet propagation phase (\sect{mag_rec}). 

Before describing the different steps of the simulations, we aim to examine the evolution of energy, focusing particularly on the assessment of each energy component within each simulation and its alignment with prior research findings. 

\begin{figure}
  \centering
\resizebox{0.5\hsize}{!}{\includegraphics{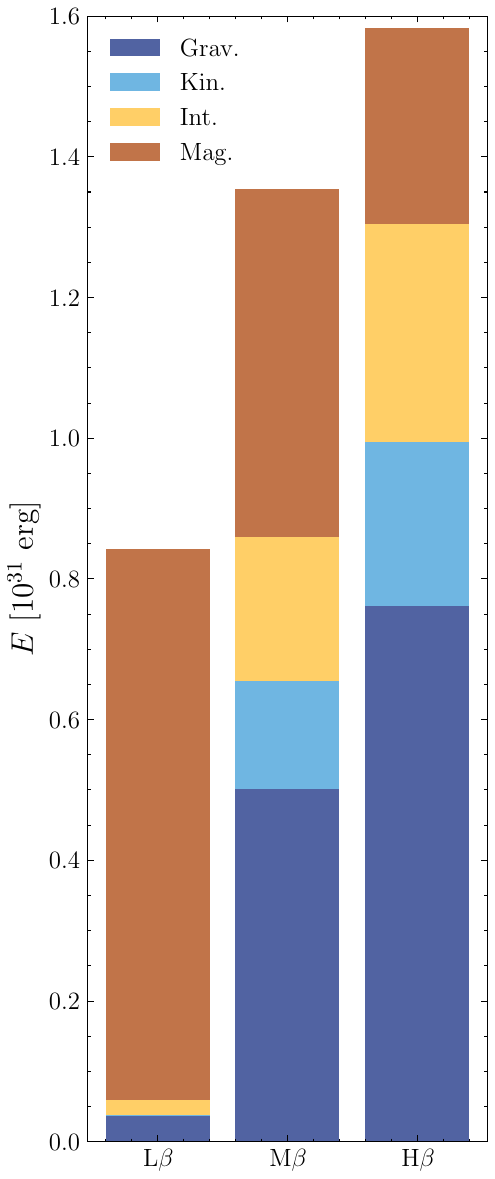}}
\caption{Stacked bar chart illustrating the distribution of energy components defined by Eqs. (\ref{eq:nrj_mag} - \ref{eq:nrj_grav}): gravitational (indigo), kinetic (blue), internal (yellow), and magnetic (orange) for the three simulations. Each bar represents the aggregated energy budget for a specific simulation, where different colours denote the contributions of the respective energy components.}
  \label{fig:energy_comparison}
\end{figure}

\Fig{energy_comparison} illustrates the distribution of energy components (gravitational, kinetic, internal, and magnetic) for each simulation, along with the combined energy budget for each scenario. The different energy components are defined as:

\begin{gather} 
    E_{\mathrm{mag}} = \int \int \int_V \frac{1}{2 \mu_0} \boldsymbol{B}^2\ dV , \label{eq:nrj_mag} \\
    E_{\mathrm{kin}} = \int \int \int_V \frac{1}{2}\rho \boldsymbol{v}^2\ dV , \label{eq:nrj_kin} \\
    E_{\mathrm{int}} = \int \int \int_V \frac{P}{\gamma -1}\ dV , \label{eq:nrj_int} \\
    E_{\mathrm{grav}} = \int \int \int_V g_{\sun} R_{\sun}^2 \rho r\ dr\ d\theta\ d\phi \, \text{.} \label{eq:nrj_grav} 
\end{gather}
The analysis of the figure reveals several noteworthy observations. Firstly, there is a discernible trend wherein the total energy sum appears to increase with higher plasma $\beta$ values. Particularly, the combined energies exhibit closer proximity in simulations with M$\beta$ to H$\beta$ values. Additionally, distinct trends emerge indicating a decrease in magnetic energy and concurrent increase in internal, gravitational, and kinetic energies as plasma $\beta$ rises. These trends align with previous findings: L$\beta$, indicative of a plasma $\beta$ parameter below 1, exhibits dynamics predominantly governed by magnetic field lines, thereby resulting in magnetic energy dominance over other energy components. In contrast, M$\beta$ and H$\beta$ demonstrate a transition towards dynamics dominated by thermodynamic effects beyond a certain solar radius. Notably, closer proximity of the plasma $\beta = 1$ surface (and the Alfvén surface) to the sun, in H$\beta$, accentuates the prominence of internal, gravitational, and kinetic energy proportions relative to M$\beta$.

\begin{figure}
  \centering
  \resizebox{0.9\hsize}{!}{\includegraphics{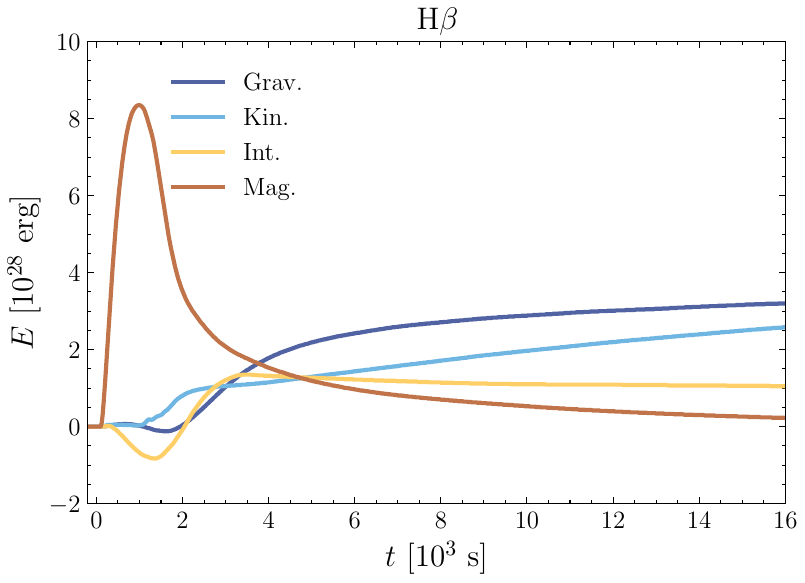} }
  \caption{Temporal evolution of gravitational (indigo), kinetic (blue), internal (yellow), and magnetic (orange) energy variations for the H$\beta$ simulation. Energies are computed relative to their values at time $t=0$: $(\mathrm{E}(t) - \mathrm{E}(t=0))$. The reference time, $t=0$, is defined as $100$ seconds before the onset of magnetic injection.
  }
  \label{fig:nrj_HB}
\end{figure}

Focusing on H$\beta$, \fig{nrj_HB} shows the temporal evolution of gravitational, kinetic, internal, and magnetic energies. These fluctuations are computed as the energies at time $t$ subtracted by the energies at the initial time ($t=0$). The reference time ($t=0$) is defined as $100$ seconds preceding the commencement of magnetic injection. This plot reveals four distinct phases: (1) a period of relaxation, (2) a gradual accumulation of energy induced by photospheric forcing, (3) a brief, explosive onset of reconnection driven by instability, resulting in energy release and the generation of a jet; and (4) an extensive propagation phase that begins with the jet generation, each of which will be delineated subsequently.

\subsubsection{Relaxation phase} \label{sec:relax}

Due to the influence of the solar wind flow, the system initially experiences an imbalance in forces. To attain a quasi-steady state where the magnetic forces and kinetic pressure of the wind counteract this initial imbalance, the system requires a relaxation phase. In our approach, we initially executed the simulation without dynamic grid refinement nor photospheric forcing, allowing the system to relax until both magnetic and kinetic energy reached a nearly constant state. The duration of the relaxation phase varies among simulations. For L$\beta$, it is only $100$ s, while for M$\beta$ it is $71\,100$ s and for H$\beta$ it is $65\,100$ s. \Fig{nrj_HB} shows a part of this phase with a period spanning from $t = -200$ s to $t = 100$ s, with the reference time, $t = 0$, corresponding to $100$ s before the magnetic injection. We can see in \fig{nrj_HB}, the energy variations are negligible compared to the variations in the later phases. 

\subsubsection{Photospheric forcing and twisting motions} \label{sec:photo}

In order to energise the system and provide the magnetic energy to the close-field system that will be fuelling the jet, a quasi-steady rotational photospheric flow was applied to the plasma on the bottom surface over the entire parasitic polarity, that is, the one of the embedded dipole. This mimics the slow shearing and rotational flows observed at the magnetic polarities on the Sun. The spatial profile of the photospheric flow, tangential velocity, $v_{\mathrm{b}}$, depends on the gradient of the radial magnetic field, $B_r$, and  is given by the following equation (\citetalias{Karpen17}):
\begin{equation}
\boldsymbol{v_\mathrm{b}} = \boldsymbol{v}\Bigl\arrowvert_{r=R_\sun} \Bigr. = v_0 f(t) \lambda_{\mathrm{b}} \frac{B_2 - B_1}{B_r} \tanh \left( \lambda_{\mathrm{b}} \frac{B_r -B_1}{B_2 - B_1}\right) \boldsymbol{\hat{r}}\times \boldsymbol{\nabla}_t B_r\Bigl\arrowvert_{r=R_\sun} \Bigr.
\end{equation}
with $v_0$ = $20$ ($25$ for H$\beta$) $\times 10^{12}\  \mathrm{cm}^2 \mathrm{s}^{-1} \mathrm{G}^{-1}$, $B_2$ and $B_1$, user defined upper and lower bounds on where the flow is applied. This yields a peak flow of $v_\mathrm{b}$ = $88$ km $\mathrm{s}^{-1}$ (respectively $59$ km.$\mathrm{s}^{-1}$ and $114$ km $\mathrm{s}^{-1}$), about  $69\,\%$ (respectively $32\,\%$ and $62\,\%$) of the coronal sound speed and only $1.3\,\%$ (respectively $0.7\,\%$ and $1.5\,\%$) of the peak Alfvén speed at the surface for the L$\beta$ (respectively M$\beta$ and H$\beta$). 

The velocity flow is applied by adding a temporal ramp profile using the cosine function below: 
\begin{equation}
    f(t) = \frac{1}{2} \left[1- \cos \left( \pi \frac{t - t_{\mathrm{max}}}{t_{\mathrm{max}} - t_{\mathrm{min}}}\right) \right]
\end{equation}
with $t \in [t_{\mathrm{min}} = 100\,$s, $t_{\mathrm{max}} = 1\,100\,$s]. $t_{\mathrm{min}}$, $t_{\mathrm{max}}$ represent the initial and final time of application of the flow after the end of the relaxation time. The imposed flow $v_\mathrm{b}$ is in the clockwise direction and imparts a right-handed twist to the field. Contrary to \citetalias{Karpen17}, the flow is smoothly decelerated to rest at $t = t_{\mathrm{max}}$, similarly to \citet{Pariat09a}. 

The fan is not directly stressed. The flow applied adheres to the contours of $B_r$, thus maintaining its surface distributions unchanged over time. Consequently, the potential field remains temporally roughly constant \citep{Linan20}, facilitating the tracking of the system's temporal evolution of free magnetic energy, by calculating the difference between the total magnetic energy and its initial value. 

During this phase, magnetic energy increases while internal energy decreases (cf. \fig{nrj_HB}). The imposed clockwise flow induces a counter-clockwise twist to the closed magnetic field lines, boosting the total field strength and reducing the pressure within the expanding separatrix dome. This behaviour explains the magnetic energy's upward trend and the internal energy's decline seen in \fig{nrj_HB}. However, variations in kinetic energy due to tangential flow and separatrix dome expansion, as well as gravitational energy, remain negligible (see \fig{nrj_HB}).

\begin{figure}
  \centering
    \resizebox{0.8\hsize}{!}{\includegraphics{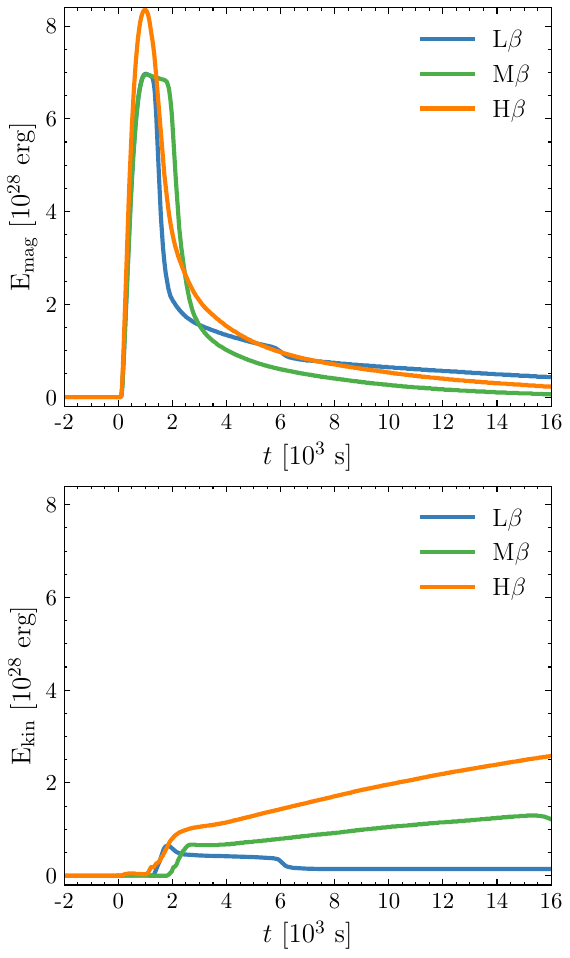}}
  \caption{Temporal evolution of magnetic (top) and kinetic (bottom) energy variations for the three simulations: L$\beta$ (blue), M$\beta$ (green), and H$\beta$ (orange). Energies are computed relative to their values at time $t=0$: $(\mathrm{E}(t) - \mathrm{E}(t=0))$. The reference time, $t=0$, is defined as $100$ seconds before the onset of magnetic injection.}
  \label{fig:mag_kin_nrj_plots}
\end{figure}

In \fig{mag_kin_nrj_plots}, the temporal evolution of the magnetic (top panel) and kinetic energy (bottom) variations
are displayed for the three simulations. In the three simulations, during the relaxation, the magnetic energies stay roughly constant. Then, during the photospheric forcing, the magnetic energy increase with the same trend for the three simulations: even though the simulations have three different atmospheres, they are consistent in terms of photospheric forcing, twist injection. Whereas $L\beta$ and $M\beta$ have the same peak value, at $7\times 10^{28}$ erg, $H\beta$ exhibits a peak at higher value: $8.4 \times 10^{28}$ erg. This difference can be explained by the difference of speed of the photospheric flow. 

\subsubsection{Magnetic reconnection and propagation of the jet} \label{sec:mag_rec}

As the energy-build-up phase concludes, the evolution suddenly transitions to an explosively dynamic phase. This phase begins with the onset of instability and the initiation of fast reconnection. Rapid conversion of magnetic energy into kinetic energy follows. The process culminates in the generation of the coronal jet. It is important to note that the jet generation process is entirely self-consistent; the amount of twist is not a parameter we control or choose.
The initially symmetrical separatrix dome undergoes asymmetry at $t = 900$ s, causing the buckling of one side of the separatrix tower. This results in the rapid flattening of the null-point, dispersion of magnetic field into a current patch, and subsequent fragmentation. This process results in the unwinding of the magnetic field lines, effectively mitigating the accumulated twist during the energy build-up, thus elucidating the abrupt decline in magnetic energy variations seen in \fig{mag_kin_nrj_plots}. Notably, this unwinding entails the reconnection of closed, twisted magnetic field lines with open magnetic field lines, facilitating the propagation of the twist towards both the outer corona and the inner heliosphere. 

Throughout this phase, the newly opened magnetic field lines, exemplified by the one depicted in pink in \fig{3D_snapshots_test}, exhibit substantial bending and entanglement, forming a 'U-loop' configuration (defined in \sect{id_jet}). The imparted twist to the reconnected open field lines initiates the generation of nonlinear Alfvén waves, which in turn drives plasma outflow through wave-induced pressure gradients. Consequently, at the onset of the instability, kinetic energy escalates (see \fig{mag_kin_nrj_plots}), marked by the immediate initiation of a wave propagating at a velocity of $100$ km $\mathrm{s}^{-1}$ for H$\beta$. Subsequently, after $1\,300$ s, heightened pressure at the apex of the former dome separatrix triggers the release of plasma outflow, reaching velocities close to $300$ km $\mathrm{s}^{-1}$, close to the local Alfvén velocity in H$\beta$ (see \fig{atmosphere_conditions_v1}).
This event initiates the upward propagation of  plasma outflows, resulting in a significant increase in pressure. As a consequence, the internal energy, previously in decline, undergoes a sudden and substantial increase (see \fig{nrj_HB}). 

In \fig{mag_kin_nrj_plots}, we can see that this phase begins with the concurrent reaching of a maximum in the magnetic energy variations and an onset of a steep increase in the kinetic energy variations for the three simulations at $t = 1\,200$ s (respectively $t = 1\,900$ s and $t = 1\,100$ s) for L$\beta$ (respectively  M$\beta$ and H$\beta$). During this phase, the decrease in magnetic energy and steep increase in kinetic energy suggest a rapid conversion of magnetic to kinetic energy and can be seen for the three simulations. 

After the magnetic reconnection, the jet is generated from the emergence of nonlinear Alfvén waves and plasma flows triggered by intense kink-driven reconnection events. This jet extends its propagation into both the outer corona and the inner heliosphere. As the wave pressure exerts its force, it compresses the plasma, thereby augmenting the density of the jet material trailing behind the Alfvénic wave front, surpassing the local ambient value of the solar wind. Progressing at the coronal Alfvén speed, which can reach up to $420$ km $\mathrm{s}^{-1}$ (\fig{comparison_velocities}), the front can cross a solar radius in about $1\,300$ s, as observed in $H\beta$.
The propagation of the wave pressure results in increased compression of the plasma, consequently causing a sharp rise in internal energy initially. Subsequently, the internal energy becomes stable (see \fig{nrj_HB}).
Furthermore, the compression of the plasma by the wave pressure enhances the density of the jet material behind the Alfvénic wave front, leading to an augmentation of gravitational energy as the denser portions of the jet propagate away from the solar surface.
Finally, as the jet progresses, its various segments accelerate, resulting in a increase in kinetic energy variations (see \fig{mag_kin_nrj_plots}).

This section demonstrates the consistent presence of distinct steps in all three simulations, indicating an auto-consistent generation of the jet. Additionally, it underscores the similarity in the injection of twisting motions across the three simulations. This injection initiates magnetic reconnection, facilitating a partial conversion of magnetic energy into kinetic energy and the jet generation and propagation. Some of the limitations of the model will be discussed in \sect{Disc}. In the next section we delve deeper into the specifics of jet propagation in the three parametric simulations, each characterised by distinct Alfvén surfaces.

\section{Comparative analysis of the dynamics of the propagating jet structures in the parametric simulations} \label{sec:jet_propagation}

In this section we analyse the dynamics of the different propagating structures identified in the parametric simulations. We begin by conducting a detailed analysis and initial identification of the substructures in the H$\beta$ case (\sect{id_jet}). Subsequently, we compare the dynamics of these structures in the three different simulations, varying according to the radial atmospheric profiles (\sect{jet_dynamics}).

\subsection{Identification of structures of the jet}  \label{sec:id_jet}

\begin{figure*}[ht!]
  \centering
 \resizebox{0.9\hsize}{!}{\includegraphics{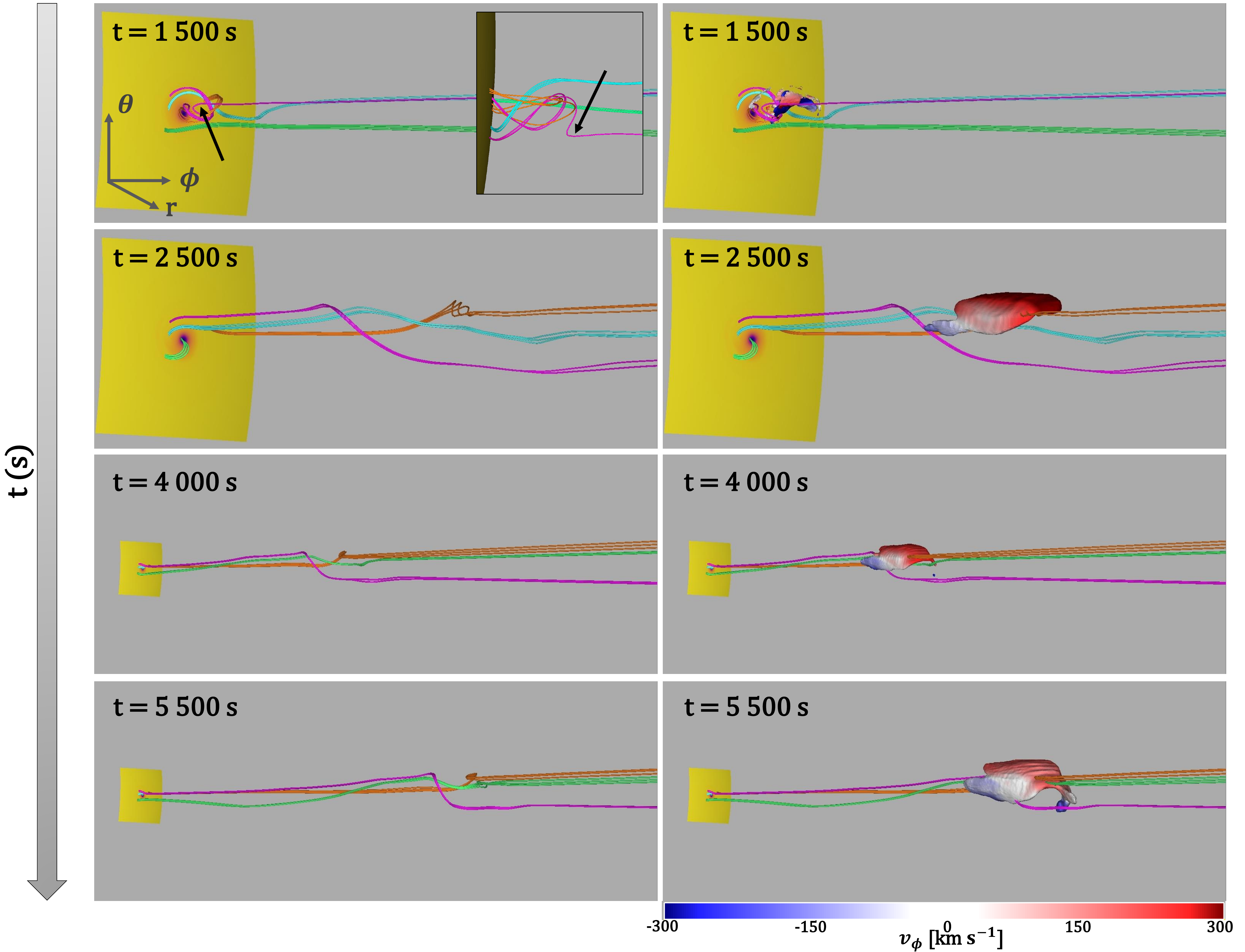}} 
\caption{Snapshots of the evolution of representative magnetic field lines and the high velocity region in H$\beta$ for four different times: $1\,500$ s, $2\,500$ s, $4\,000$ s, and $5\,500$ s. The colour-coded group of magnetic field lines are plotted from fixed points along the bottom boundary from areas where no flow was applied. The bottom boundary corresponding to the solar surface has a size of $18\degr$ in both $\theta$ and $\phi$. In the right column, the isosurface of constant velocity ($v = 300$ km $\mathrm{s}^{-1}$) is colour-coded according to the $\phi$ component of the velocity $v_{\phi}$: blue indicates out-of-plane velocity and red into-plane velocity. Zoomed images are provided for the first two time intervals ($1\,500$ s and $2\,500$ s). In the zoomed images, the magnetic fieldlines are visible up to $1.8 \, R_{\sun}$ above the solar surface whereas in the bottom pictures, the magnetic fieldlines are represented up to $4.6 R_{\sun}$ above the solar surface. An animation of this figure is available online.}
  \label{fig:3D_snapshots_test}
  \end{figure*}
  
In this section we first focus on the specific case of H$\beta$, to analyse the jet propagation and dynamics. \Fig{3D_snapshots_test} presents snapshots of the 3D jet propagation at different times with the representation of magnetic field lines and a high velocity region.   

We first examine the magnetic field lines in the left columns of \fig{3D_snapshots_test}. At $t = 1\,500$ s, which is $400$ s after the onset of kinetic energy increase (see \fig{nrj_HB}), both the cyan and green field lines are open. The formation and propagation of the untwisting magnetic wave has just started. The green lines appear relatively straight, indicating they have not yet reconnected. They represent the background magnetic environment. In contrast, the cyan lines show some bending, suggesting they reconnected slightly earlier, resulting in their opening. The bending is the signature of a propagating untwisting magnetic wave. The orange lines and most of the pink ones are closed and exhibit significant twisting. The closed pink lines have a section in closer proximity, and orthogonal with a segment of the green lines, in the vicinity of the parasitic polarity. This indicates that they are about to interchange reconnect with each other. One pink line is open, having recently reconnected, and shows pronounced twisting at its lower section. This open pink line also features a 'U-loop' shape segment. A 'U-loop' is a segment of a magnetic field line that forms a U-shape, where the bottom of the U is parallel to the surface. It can be defined as a magnetic deflection greater than $90 \degr$. This configuration can be clearly observed in the zoomed view on the top left corner of Fig. \ref{fig:3D_snapshots_test}, with the open pink magnetic field line.

At $t = 2\,500$ s, the shape of the field lines and the connectivity of some of them have changed, the latter due to ongoing magnetic reconnection. The previously open green lines are now closed, displaying moderate twist. They have undergone interchange reconnection with the orange and pink lines, which are now open. All groups of open lines exhibit a bent section. The cyan and pink lines show moderate twisting, with the bent section of the cyan lines now present further up (1.6 $R_{\sun}$). This shift is the main evidence of the propagation of an untwisting wave, as discussed in \citet{Pariat09a, Pariat16}. The newly open orange lines also display a twisted section. Although there appears to be a U-shaped segment in the orange lines, this is merely a projection effect. Other viewpoints (not shown here) confirm that while the orange field lines are twisted, they do not present any magnetic field inversion this time.

At $t = 4\,000$ s, further magnetic reconnection is observed: the previously open cyan lines are now closed, while the previously closed green lines are now open. The magnetic deflections along the orange, green, and pink field lines continue to move upwards, signalling the ongoing propagation of an untwisting wave.

At $t= 5\,500$ s, such propagation is still ongoing. The bent segment of the orange, green, and pink lines have moved further up, as the result of the ongoing radial propagation of an untwisting wave along their lengths. The orange and green field lines appear as twisted but still do not exhibit U-shaped segments. The bend of the pink lines seems increasingly marked, suggesting that the magnetic deflection is steepening as the magnetic wave front propagates upwards. 

The velocity isosurface displayed in the right column of \fig{3D_snapshots_test} reveals that the jet exhibits a horizontally rotating structure propagating through the domain over time. A careful examination reveals that the regions with twisted and bent magnetic field lines do correspond to areas of strong transverse velocities. This indicates that the rotating motion of the high-velocity region is associated with the propagation of twisted and bent structures of the magnetic field lines. Therefore, the magnetic deflection of the field lines is closely associated with a velocity enhancement, highlighting the Alfvénic nature of the wave. 

Regarding the presence of magnetic inversion, i.e. the presence of U-shaped segment along the field lines, as previously noted, it is observed that immediately after the interchange reconnection of the strongly twisted field lines, the newly formed open field lines display a U-shaped segment in the vicinity of the magnetic polarity (as with the open pink line at $t = 1\,500$ s). However, this magnetic inversion disappears rapidly as the untwisting wave propagates upwards. For example, by $t = 2\,500$ s, the pink lines no longer present any 'U-loop'. This observation holds true for all groups of field lines, with none exhibiting U-shaped segments higher than $1.11 \, R_{\sun}$. This enables us to state that a 'U-loop' does not survive in the corona.

As previously discussed in \citet{Pariat09a, Pariat15a}, \fig{3D_snapshots_test} illustrates that the jet generation is a continuous process characterised by the sequential reconnection of magnetic field lines. The newly formed magnetic field lines propagate untwisting torsional waves upwards. While U-shaped loops, indicative of magnetic field inversion, are commonly noted at the coronal base, at the jet source, they quickly disappear as the torsional wave ascends. 

\begin{figure*}
 \sidecaption
 \includegraphics[width=12cm]{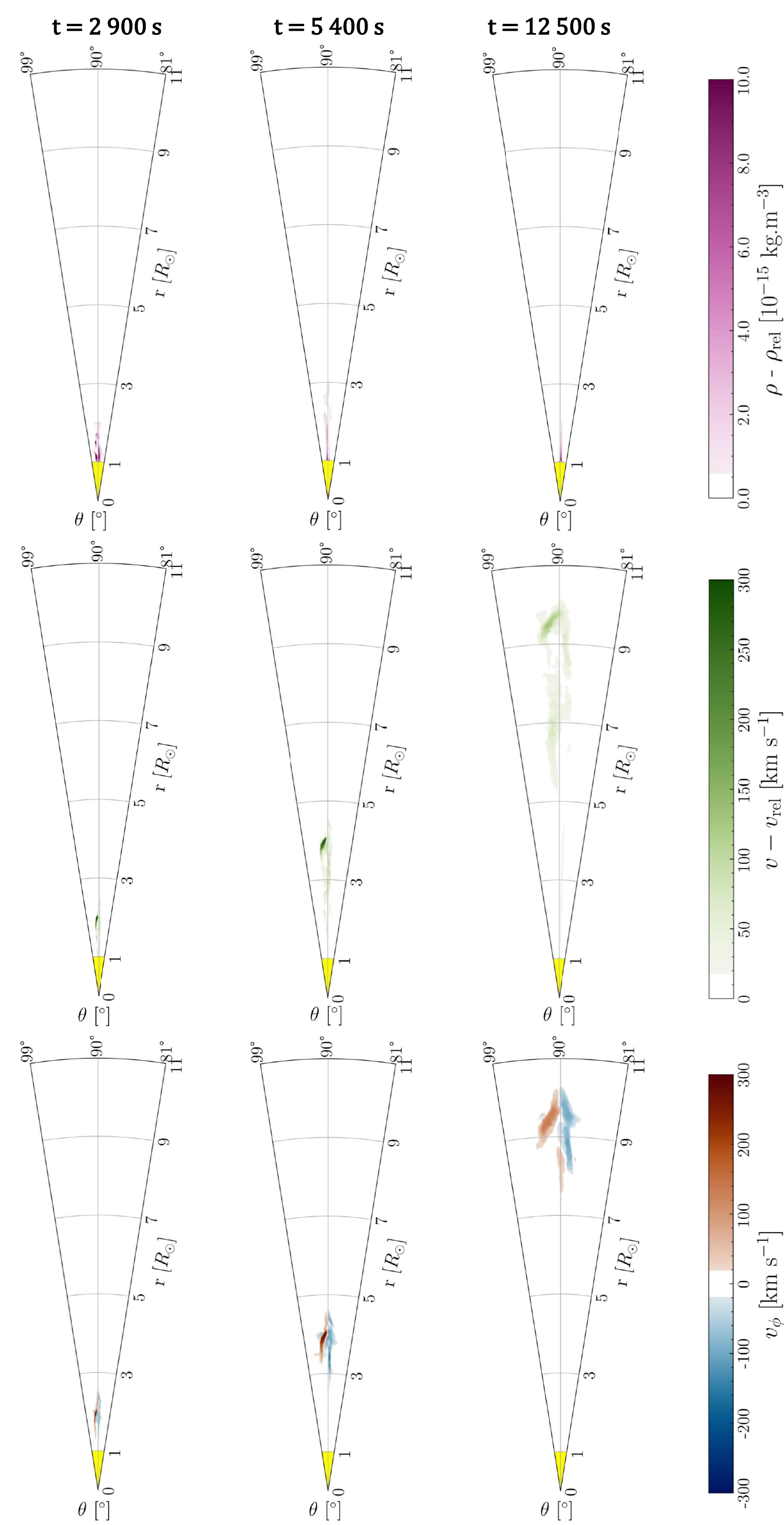}
 \caption{2D cuts at $\phi = 0\degr$ of the evolution of selected quantities in  H$\beta$ at $t = 2\,900$ s, $t = 5\,400$ s, and $t = 12\,500$ s: mass density variations, $\rho - \rho_{rel}$ (top row panels), velocity variations $v - v_{rel}$ (middle row panels), and $\phi$ velocity component, $v_\phi$ (bottom row panels), perpendicular to the plane of the cut with blue (respectively red) colour scales indicating observer-directed  (respectively oppositely directed) velocities. An animation of this figure is available online.}
  \label{fig:2Dcuts_evolution}
\end{figure*}

To deepen our analysis, we investigate the temporal evolution of various variables as the jet propagates. Given the radial propagation of the jet, we perform a cut along a specific angle within the domain: $\phi = 0\degr$. The 2D cuts at this angle allow us to examine the temporal evolution of the variables, as displayed in \fig{2Dcuts_evolution}. To clearly distinguish the variation of the physical quantities associated with the jet propagation, the background was removed by subtracting the initial values of the quantities (at $t = 100$ s, i.e. base difference).

The base difference is defined as the difference between a quantity at a given time and the same quantity at a previous reference time. In this subsection, the reference time is the end of the relaxation period, $t = 100$ s. Mathematically, the base difference for a quantity $q$ at time $t$ and position $(r, \theta, \phi)$ is given by:
\begin{equation} \label{eq:base_diff}
    \delta q (r, \theta, \phi, t) = q (r, \theta, \phi, t) - q_{rel} \,, \forall \theta \in [81\degr, 99\degr]
\end{equation}
 where $q_{rel} \equiv q(r, \theta, \phi, t=100 \,  \mathrm{s}$). As explained in \sect{param_sim}, the atmosphere is stratified, resulting in increasing background velocities and decreasing mass density with radial distance (cf. \fig{atmosphere_conditions_v1}). During jet propagation, the apparent velocity of a structure may become comparable to the surrounding environment, preventing the structure from being seen. Hence, removing the background is essential for visualising the different structures. This method is analogous to base-difference imaging in observational data, as used in \citet{Kumar19} article. We choose a cut at constant $\phi = 0 \degr$ because the centre of the jet is approximately at this location. 

\Fig{2Dcuts_evolution} presents the spatial distribution of the mass density variations, $\delta \rho = \rho(t) - \rho_{rel}$, the velocity variations, $\delta v = v(t) - v_{rel}$, and the $\phi$ component of the velocity, $v_{\phi}$. 
The mass density variations, represented in the top panel of \fig{2Dcuts_evolution}, show that there is a dense structure that extends through the domain. Indeed, at $t = 2\, 900$ s, a dense structure (plotted in darker red) is present from the solar surface to about 1.97 $R_{\sun}$, whereas at $t = 5\, 400$ s, the structure is extending further to nearly $3 \, R_{\sun}$. At $t = 12\,500$ s, the dense structure seems to extend only to $2.2 R_{\sun}$ with the colour scale that was selected. However, at $t = 18\,000$ s, the dense structure extends to $11.5 \, R_{\sun}$ with a density of $\delta \rho = 7 \times 10^{-17} \, \mathrm{kg} \, \mathrm{m}^{-3}$. 
Moreover, the mass density seems to decrease temporally close to the solar surface: $\rho (r = 1.27 \, R_{\sun}, \, \theta = 89.97 \degr, \, \phi = 0 \degr, \, t = 5\,400 \, \mathrm{s}) - \rho (r = 1.27 \, R_{\sun}, \, \theta = 89.97 \degr, \, \phi = 0 \degr, \, t = 2\,900 \, \mathrm{s}) = 8.5\times10^{-15} \, \mathrm{kg}\,\mathrm{m}^{-3}$. Thus, the flow of mass density seems to scatter through the simulation domain.
In addition to that, it seems that there is a launch of denser plasma at $t = 12\, 500$ s: there is a local temporal increase close to the solar surface, almost $4 \times 10^{-15} \mathrm{kg}\,\mathrm{m}^{-3}$ between $t = 5\,400$ s and $t = 12\, 500$ s. 

In the middle panels in \fig{2Dcuts_evolution}, the variations in velocity magnitude are depicted. We can observe two different structures. At $t = 2\,900$ s, a region with a high increment of velocity (with a  maximum value of $\delta v = 357 \,  \mathrm{km}\,\mathrm{s}^{-1}$) is observed at $1.97 \, R_{\sun}$, preceded, at a smaller radius,  by areas with lower velocity increment ($\delta v \sim 120 \, \mathrm{km}\,\mathrm{s}^{-1}$). This pattern persists for the two other time steps. Indeed, at $t = 5\, 400$ s, there is a region of higher velocity with a maximum $\delta v = 301\, \mathrm{km}\, \mathrm{s}^{-1}$ near $4R_{\sun}$ followed by regions of lower $\delta v$, around $80\, \mathrm{km}\, \mathrm{s}^{-1}$.  At $t = 12\, 500$ s, the region of higher velocity with a maximum $\delta v = 108\, \mathrm{km}\, \mathrm{s}^{-1}$ is located at $9.42\,R_{\sun}$ followed by regions of lower $\delta v$, around $50\, \mathrm{km}\, \mathrm{s}^{-1}$. In addition to that, we can see that the prominent structure with the higher $\delta v$ has a significant temporal decrease: at $t = 5\, 400$ s, the maximum $\delta v$ is $301\, \mathrm{km}\, \mathrm{s}^{-1}$, whereas at $t = 12\, 500$ s, the maximum $\delta v$ is $108\, \mathrm{km}\, \mathrm{s}^{-1}$. The other areas with lower velocity increment have decreased almost as markedly: a temporal decrease of around $50\%$ between  $t = 5\,400$ s and $t = 12\,500$ s. 
By comparing the top and middle panels, we can notice that the areas with a small increase in velocity correspond to areas with an increase in mass density but the high velocity region is on the other hand not correlated with a region of sensitive mass density increment. This will be investigated more in depth in the following paragraphs (see \fig{wave_front}, \fig{RT_diagramm}).

The $\phi$ component of the velocity highlights the rotational motions of the jet. The blue component is predominantly located below $\theta = 90\degr$, with a small patch above this angle. Conversely, the red component is primarily situated above $\theta = 90\degr$. Despite their propagation, these patches appear to maintain roughly the same angle locations: the higher positive $\phi$ velocity component is located above the probe $\theta = 90 \degr$ while the lower negative $\phi$ velocity component is located below the probe $\theta = 90\degr$. At the latest time step ($t =12\,400$ s), a new patch of positive velocity appears
close to $\theta = 90\degr$.

When comparing the $\phi$ component of the velocity and velocity structures, we observe that the leading edge of the high-velocity structure aligns with regions exhibiting the highest rotational velocities. These rotational structures do not appear to correspond with variations in mass density. However, let us note that regions with slower rotational velocities (less than $25$ km $\mathrm{s}^{-1}$) are associated with changes in mass density (not discernible in \fig{2Dcuts_evolution} because of the choice of the range of values).

\Fig{2Dcuts_evolution} suggests the presence of multiple distinct structures within the jet, which can be categorised into two types:
\begin{itemize}
    \item a high velocity, very low mass density structure: this region exhibits high velocity increase without a corresponding significant increase in mass density, and associated with the jet's rotational component. This pattern may indicate the presence of (near-) incompressible waves. It appears to correspond to areas observed in \fig{3D_snapshots_test}, where a portion of the isosurface is linked with a high-velocity rotation and a deflection in the magnetic field lines.
    \item high-mass density, low-velocity structures: these regions have higher mass density but lower velocity, with little or no rotating motions, suggesting a dense bulk flow of plasma.
\end{itemize}
These results are consistent with findings in \citet{Pariat16, Uritsky17, Roberts18}. To assess the consistency of these structures across different simulations, we will now investigate if similar substructures are present in the other two simulations.

\subsection{Comparative analysis of jet dynamics in three different parametric simulations}\label{sec:jet_dynamics}

\begin{figure*}
  \centering
 \resizebox{0.9\hsize}{!}{\includegraphics{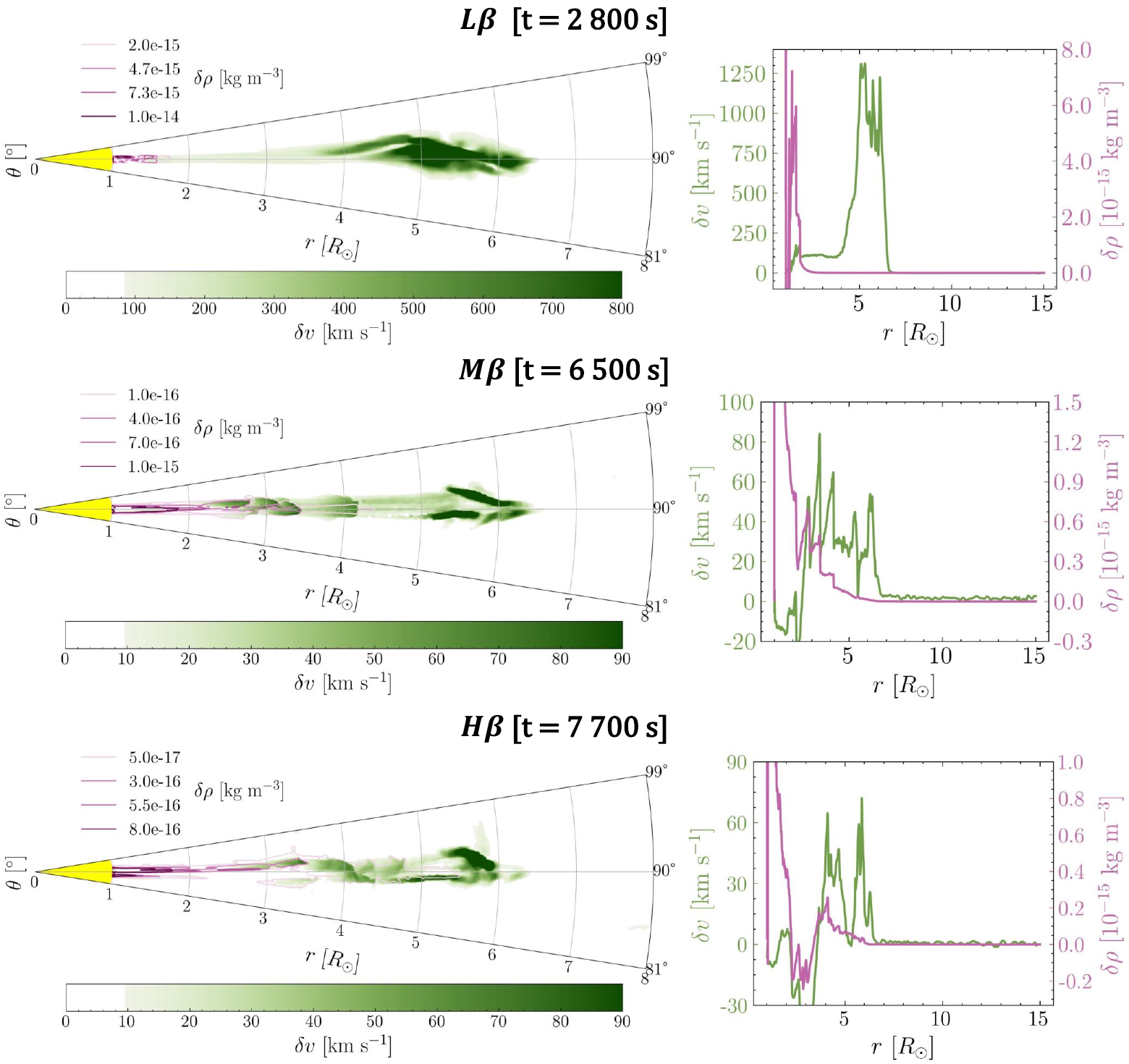}
 }
  \caption{2D distribution (at $\phi = 0\degr$) of the velocity variations, $\delta v = v - v_{\mathrm{rel}}$, in green with isocontours for mass density variations, $\delta \rho = \rho - \rho_{\mathrm{rel}}$, in red. The isocontours for mass density variations range from $2\times 10^{-15}$ to $10^{-14}$ kg $\mathrm{m}^{-3}$ (respectively $10^{-16}$ to $10^{-15}$ kg $\mathrm{m}^{-3}$ and $5 \times 10^{-17}$ to $8 \times 10^{-16}$ kg $\mathrm{m}^{-3}$) in L$\beta$ (respectively M$\beta$ and H$\beta$). On the right part, radial probes (at $\phi = 0\degr$ and $\theta = 90\degr$) are displayed for both quantities. An animation of this figure is available online.}
  \label{fig:wave_front}
\end{figure*}

In this section we examine the three parametric simulations to determine if the previously observed structures are present in the three simulations. We focus on two quantities: mass density, $\delta \rho$, and velocity, $\delta v$, variations, by removing the background radial profile (see \eq{base_diff}). In \fig{wave_front}, we compare the spatial distribution of mass density and velocity in order to determine the relative location of the high velocity region and the dense flow region in each simulation. We analyse each simulation when the front of the high velocity structure has reached a given radius of $6.3 \, R_{\sun}$. In each simulation, this radius is specifically reached at different instants: $t = 2\,800$ s (respectively $t = 6\,500$ s and $t = 7\,700$ s) for L$\beta$ (respectively M$\beta$ and H$\beta$). This variation in timing is due to the differing speeds at which the front travels, influenced by the properties of the atmospheric stratification. This indicates that the high velocity structure propagates significantly faster in L$\beta$ and slightly faster in M$\beta$ compared to H$\beta$.  To the right of the spatial distribution of mass density and velocity variations, we can see radial probes of these two quantities. In this figure, a radial probe corresponds to the radial distribution of the given quantities along the central spine ($\theta = 90\degr$ and $\phi = 0\degr$). The radial probe effectively reveals the variations in each quantity and their fronts. 

We can note in L$\beta$ simulation of \fig{wave_front}, that the structure of increased mass density is primarily located between $1$ and $1.8 \, R_{\sun}$, whereas the high velocity structure extends from $4 \, R_{\sun}$ to $6.3 \, R_{\sun}$. Below $4 \, R_{\sun}$, the increase in plasma velocity is lower than $100$ km $\mathrm{s}^{-1}$, while it is above $800$ km $\mathrm{s}^{-1}$ in the high velocity front. The radial probe shows that the mass density is higher below $2 \, R_{\sun}$, location where the velocity enhancement is only moderate. Moreover, between $1.8 \, R_{\sun}$ and $4 \, R_{sun}$, there is a region with minimal mass density and a nearly constant velocity around 100 km $\mathrm{s}^{-1}$. The previously identified substructures are present: a high velocity structure with no increase in mass density and a bulk flow of plasma between $1$ and $1.8 \, R_{\sun}$. 

In the M$\beta$ simulation (middle panel of \fig{wave_front}), mass density variations are an order of magnitude lower than in L$\beta$, spanning from $10^{-16}$ to $10^{-15} \, \mathrm{kg}\, \mathrm{m}^{-3}$. The high density region reaches $4.3 \, R_{\sun}$ but thanks to the isocontours, we can observe that the relative mass overdensity decreases radially. The leading edge of the high velocity region has a velocity around $300$ km $\mathrm{s}^{-1}$ and is not associated with a significant relative increase in mass density. We can observe that in M$\beta$, unlike in L$\beta$, the high velocity region is not located on the central spine (i.e. at $\theta = 90\degr$ and $\phi = 0\degr$), but surrounds it, having a 3D helical shape. In the right part of the middle panel, the radial probe shows, close to the Sun, a high mass density structure with a negative velocity variation relative to the background. Then, between $2.3$ and $4.3 \, R_{\sun}$, there are three mass density variations peaks that coincide with velocity variations peaks. Above $4.3 R_{\sun}$, there is the high velocity structure that coincide with a really low mass density variations region. 

For the H$\beta$ case (lower panel of \fig{wave_front}), the velocity increase relative to the background atmosphere can reach up to $214$ km $\mathrm{s}^{-1}$ which is close to the values of M$\beta$. However, the relative mass density increase is about one magnitude lower. The high density region is close to the solar surface and this sensitive increase in mass density can be noted up to $5.5 R_{\sun}$ (faintest isosurface). Some mass density variations isocontours match patches of velocity between $3$ and up to $5.5 R_{\sun}$. The leading edge of the high velocity enhancement structure at $6.3 R_{\sun}$ has lower velocities than the previous simulations. Similarly to the two previous $\beta$ cases, the front of this high velocity region is not associated with a noticeable increase in mass density. Moreover, the high velocity region is again not centred on the spine, but rather adopt a helical shape around the central axis. 

Thus, the two substructures identified in H$\beta$, that are the high velocity front with no mass density enhancement, and the high density region associated with moderate velocity, are also present in the other two simulations. With increasing plasma $\beta$ between the simulations, the spatial separation between structures decreases.

\begin{figure*}
  \centering
\includegraphics[width=17cm]{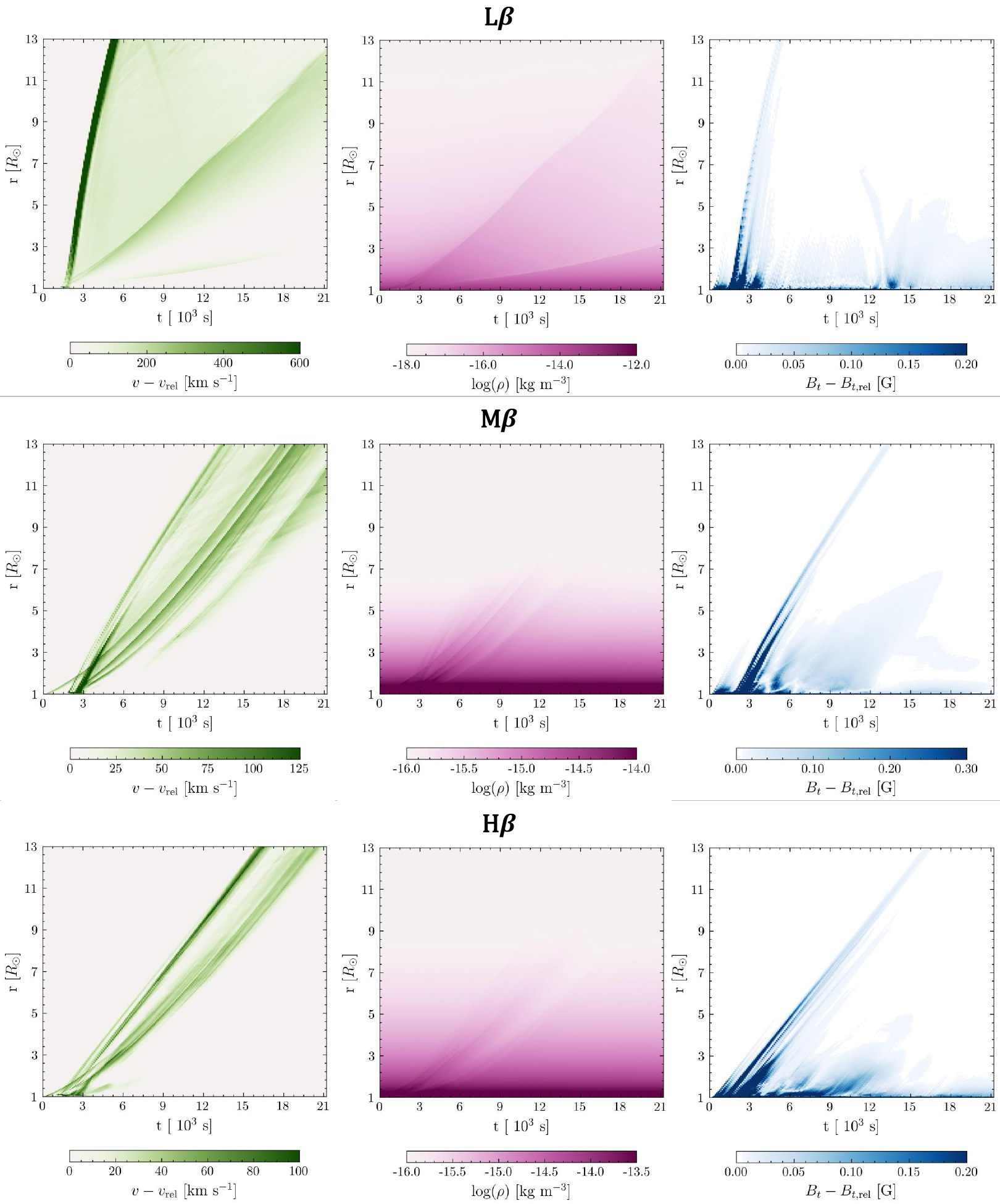}
  \caption{Radius-time diagrams, at $\theta = 90\degr$ and $\phi = 0\degr$, of the velocity variations $v - v_{rel}$,  the logarithm of the mass density, and the variations of the transverse component of the magnetic field, $B_t -B_{t,rel}$, with $B_t = \sqrt{B_{\theta}^2 + B_{\phi}^2}$ for the three different simulations.}
  \label{fig:RT_diagramm}
\end{figure*}

In order to analyse more precisely the propagation speed of the different structures, \fig{RT_diagramm} presents radius-time (r-t) diagrams of the base difference (see \eq{base_diff}) of the velocity, the logarithm of mass density and the base difference of the transverse magnetic field (being the square root of the sum of squared $\theta$ and $\phi$ components), for each parametric simulation. These diagrams were built by taking the radial profiles of the quantities along the central spine of the domain (at $\theta = 90 \degr$ and $\phi = 0 \degr$) for each output of the simulations. This approach means that one radial probe at a given time corresponds to a vertical line in one plot in \fig{RT_diagramm}. These kinds of plots enable the propagation of structures to be seen and for their apparent propagation speed to be estimated directly thanks to the slope of the patterns present in the r-t diagram. This method is commonly used in observations \citep[e.g.][]{Sheeley99,Morton12, Schmieder13}. 

The top panel of \fig{RT_diagramm} corresponds to L$\beta$. The first diagram displays velocity variations and reveals a first structure with a high velocity increase. The mean value of the velocity increase (above the solar wind) is $\delta v \sim 1\,600$ km $\mathrm{s}^{-1}$. The mean linear speed estimated from the slope indicates an apparent radial propagation speed about $2\,300$ km $\mathrm{s}^{-1}$. In contrast, the mean value of the radial velocity yields $v_r \sim 1\,200$ km $\, \mathrm{s}^{-1}$.  The apparent propagation speed is thus significantly larger than the actual radial velocity of the plasma, revealing the wave nature of this leading propagation pattern. The slope in the r-t diagram thus corresponds to the phase speed of the wave and the actual plasma velocity the group speed.
This pattern does not appear in the plot of the logarithm of the mass density, suggesting it is thus not associated with a significant density increase. The wave is thus incompressible or weakly compressible. Conversely, this pattern is clearly discernible in the transverse magnetic field increase, highlighting its magnetic nature. All these elements thus suggest that this leading high-propagation-velocity structure is an Alfvénic wave. This structure is the torsional wave induced by the magnetic untwisting mechanism of the jet, as previously discussed by \citet{Pariat16, Uritsky17, Roberts18}.  

In the plot of the velocity variations, we can see a second propagation pattern, having a less inclined slope (about $400$ km $ \mathrm{s}^{-1}$). This pattern is associated with a weaker plasma velocity increase (above the solar wind), with a mean value of $\delta v \sim 270$ km $ \mathrm{s}^{-1}$. The plasma radial velocity, $v_r$, of this pattern ranges between  $190$ and $550$ km $\mathrm{s}^{-1}$ , and with a mean value of $370$ km $\mathrm{s}^{-1}$. The real radial plasma velocity and the apparent radial propagation speed are thus corresponding. The velocity is mainly radial with no significant transverse component of the velocity field (not shown here). This propagation pattern appears also in the logarithm of mass density plot, with the same slope (hence same speed), but not in the transverse magnetic field. Thus, we can deduce that this structure is a bulk flow of plasma. This structure was previously considered in \citet{Pariat16, Uritsky17, Roberts18}.  

Moreover, in the plot of the velocity variations, we can see a third structure with a low velocity variation: a mean value around  $120$ km $\mathrm{s}^{-1}$ above the solar wind and a slope of $80$ km $\mathrm{s}^{-1}$. This structure appears also in the logarithm of mass density plot, but not in the transverse magnetic field. 

For the other two simulations displayed in the lower panels of \fig{RT_diagramm}, the same structures are present:
\begin{itemize}
    \item a propagation pattern corresponding to a high velocity structure, having a mean velocity increase, $\delta v$, higher than $130$ km $\mathrm{s}^{-1}$ above the solar wind (respectively $80$ km $\mathrm{s}^{-1}$) for M$\beta$ (respectively H$\beta$), with radial values of the plasma velocity that range between $140$ (respectively $130$) and $480$ (respectively $530$) km $\mathrm{s}^{-1}$ . On the r-t diagram, the apparent propagation speed is given by the slope, which value is about $780$ km $\mathrm{s}^{-1}$ (respectively $580$ km $\mathrm{s}^{-1}$). Moreover, the slope gives an apparent propagation speed close to the Alfvén velocity. This structure is not discernible in the logarithm of the mass density but this structure is also present in the transverse magnetic field increase plot. Thus, this propagation pattern is the signature of a torsional Alfvénic wave induced by the magnetic untwisting mechanism of the jet. 
    \item multiple structures characterised by a velocity variation, $\delta v$, of : $80$ km $\mathrm{s}^{-1}$ (respectively $70$ km $\mathrm{s}^{-1}$) for M$\beta$ (respectively H$\beta$) above the solar wind. The plasma radial velocity, $v_r$, of this pattern ranges between $140$ (respectively $130$)  and $520$ km $\mathrm{s}^{-1}$ for both simulations, and with a mean value of $480$ (respectively $475$) km $\mathrm{s}^{-1}$. These structures have a slope giving an apparent propagation velocity of $510$ km $\mathrm{s}^{-1}$ (respectively $515$ km $\mathrm{s}^{-1}$). These propagation patterns appear also in the logarithm of mass density plot with the same slope, but not in the transverse magnetic field. Hence, we can deduce that these patterns are induced by bulk flows of propagating plasma.
\end{itemize}

Therefore, in the three simulations there is a leading wavefront, propagating at Alfvénic velocity, near incompressible and associated with intense transverse velocity and magnetic field. This part is the torsional wave induced by the magnetic untwisting of the jet, as previously discussed by \citet{Pariat16, Uritsky17, Roberts18}. The other structures are bulk flows of plasma, associated with an increased mass density and mainly radially directed with almost no transverse velocity. These findings are consistent with the structures discussed in \citet{Pariat16, Uritsky17, Roberts18}. 

However, among the three simulations, there are some differences: the phase speeds of the wave front decreases for increasing plasma $\beta$ simulations while the velocity variations of the dense bulk flow of plasma increase for increasing plasma $\beta$ simulations. The speeds of these structures are influenced by the atmospheric characteristics. Thus, it leads to a propagation ratio of two structures mentioned above decreasing closer to one for increasing plasma $\beta$ simulations.

\section{Switchback in situ signature of a magnetic untwisting jet}\label{sec:mag_deflections}

Although the U-shaped loops present at the onset of the jet quickly straighten in the lower solar corona (cf. \sect{id_jet}), the Alfvénic magnetic deflection continues to propagate radially. Notably, the front part of the jet consists of a torsional Alfvénic wave packet induced by the jet's magnetic untwisting mechanism, characterised by high rotational speed and transverse magnetic field. This wave front is a potential candidate for inducing typical switchback magnetic deflection signatures when measured in situ.  This prompts several key questions: Can the leading part of the jet provide a signature consistent with a switchback ? Furthermore, how does the jet front's torsional magnetic wave evolves over time? 

\begin{figure*}
  \centering
 \resizebox{0.9\hsize}{!}{\includegraphics{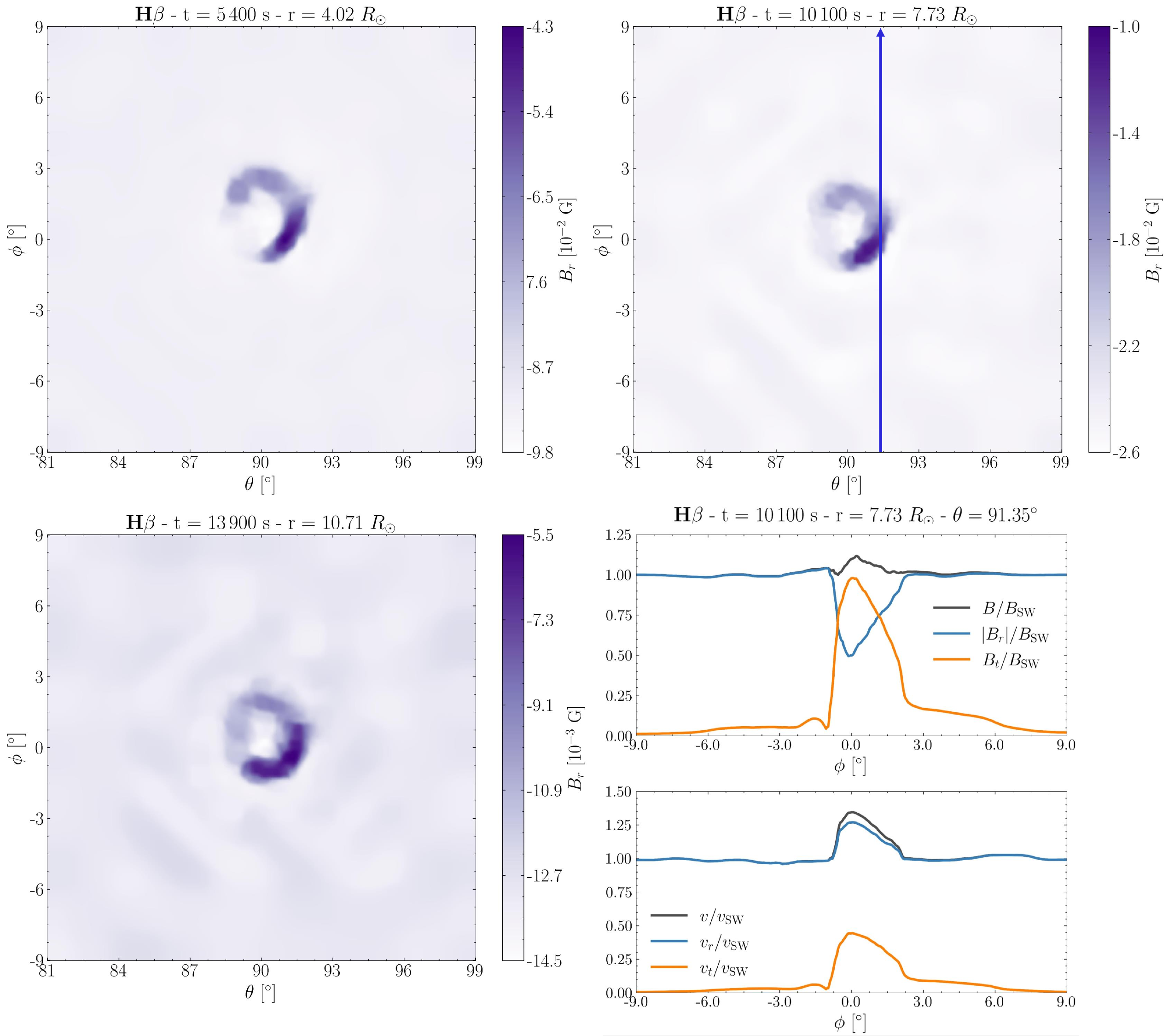} }
  \caption{Distribution at constant radius of the radial magnetic field, $B_r$, in the vicinity of the front of the jet torsional magnetic wave, at three different times: $t = 5\,400$ s (top left panel), $t = 10\,100$ s (top right panel), and $t = 13\,900$ s (bottom left panel) for the H$\beta$ simulation. For the one-dimensional probe delineated in blue in the 2D radial magnetic field cut at $t = 10\,100$ s, synthetic data for various quantities B, $B_r$, $B_t$, and V, $V_r$, $V_t$ are presented in two separate graphs (bottom right panels).}
  \label{fig:mag_deflections}
\end{figure*}

\Fig{mag_deflections} illustrates the transverse distribution (i.e. at a given radius) of the radial component of the magnetic field, $B_r$, in the vicinity of the leading edge of the torsional Alfvén wave packet at three different instants during its propagation. At $t = 5\,400$ s (respectively at  $t= 10\,100$ s and $13\,900$ s), the wave front is studied at a radius of $r = 4.02 \,R_{\sun}$ (respectively $r = 7.73 \, R_{\sun}$ and $r = 10.71 \,R_{\sun}$), where the background plasma $\beta$ is equal to $0.76$ (respectively $1.94$ and $3.45$) and the background Alfvénic Mach number, $v_{\mathrm{SW}}/v_{\mathrm{A}}$, is $0.84$  (respectively $1.95$ and $2.88$). Thus, we study the wave front's evolution, below, near and above the transition from low-to-high $\beta$ and the Alfvén surface.  

The colourmap values differed for each radial cut, showing a consistent decrease in the overall intensity of the radial magnetic field, aligned with its definition (cf. \sect{param_sim} and \fig{atmosphere_conditions_v1}). At the three time steps, a similar distribution pattern of the radial magnetic field, $B_r$, is present. A dark-purple ring like-shape structure of weaker $|B_r|$ is present inside  $\theta \in [87 \degr, 93 \degr], \phi \in [-2 \degr, 3\degr ]$. There, the magnetic field has in average 49\% (respectively 48\% and 42\%) of the background radial magnetic field intensity at $5\,400$ s (respectively $10\,100$ s and $13\,900$ s).
In particular, the weakest radial magnetic field intensity can be found at around $\theta = 92 \degr$ and $\phi = 0 \degr$ for the three instants. However, there are some differences for the distribution of the radial magnetic field for the three instants: the structure narrows after the plasma $\beta$ equals one, with visible turbulent cells forming in the distribution : darker cells that appear in the background radial magnetic field meaning a weaker radial magnetic field intensity, still higher than the one of the darkest purple structure. 

At $t=10\,100$ s, we produce synthetic in situ like measurements akin to those of the PSP instrument of the FIELDS \citep{Bale16} and Solar Wind Electrons Alphas and Protons (SWEAP) \citep{Kasper16} suites. The lower right panels of \fig{mag_deflections}  present the co-latitudinal profiles (at $r = 7.73 \, R_{\sun}$ and $\theta = 91.35\degr$) of the intensity, radial and transverse components of both the magnetic and velocity fields, respectively normalised by the ambient solar wind values of the wind speed, $v_{\mathrm{SW}}$, and the magnetic field intensity, $B_{\mathrm{SW}}$. The later quantities are obtained by taking the mean of the solar wind speed and the magnetic field intensity in the environment, here for $\phi$ in [$-9\degr$, $-3\degr$]$\cup$[$3\degr$, $9\degr$]:
\begin{eqnarray}
    q_{\mathrm{SW}} (r, \theta, t) &=& \overline{q}(r, \theta, \phi \in [-9\degr, -3\degr]\cup[3\degr, 9\degr], t) \label{eq:eq_sw_theta} \\
    q_{\mathrm{SW}} (r, \phi, t) &=& \overline{q}(r, \theta\in [81\degr, 87\degr]\cup[93\degr, 99\degr], \phi,  t) \label{eq:eq_sw_phi}
.\end{eqnarray}
The spacecraft trajectories considered in our study assume an idealised probe with unrealistically rapid motion and simplified path characteristics. Specifically, we consider that the spacecraft has infinite speed, allowing it to sample a region spanning $2.43 \, R_{\sun} = 1.7 \times 10^9$ m in just one second. Additionally, the trajectory is constrained to vary along only one angle while the other remains constant, simplifying the initially three-dimensional motion to a one dimensional motion along either $\theta$ or $\phi$ angles. This abstraction, although not reflective of actual spacecraft trajectories, enables us to isolate and analyse specific dynamical phenomena without the complexities introduced by realistic motion and temporal limitations.

The synthetic profiles of \fig{mag_deflections} show no inversion (change of sign) of the magnetic field, as already found earlier. Despite no full reversal, the absolute value of the normalised radial magnetic field drops from $1$ to about $0.5$ within $1\degr$ in $\phi$, then increases back up. The normalised transverse magnetic field concurrently rises from 0 to nearly 1, indicating that the magnetic vector evolves from purely radial to predominantly horizontal. During this magnetic deflection, the value of the norm of the magnetic field undergoes small variations, with the maximum value of $|B|/|B_{\mathrm{SW}}|$ is: $1.12$. This is less than $15 \%$ of the magnetic field norm of the background solar wind between $\phi = -0.6 \degr$ and $1.5 \degr$. This is consistent with the observations of SB (e.g. \citet{Larosa21}). Velocity measurements indicate a $40 \% $ increase in the radial velocity from the ambient solar wind radial velocity, $v_{\mathrm{SW}}$, simultaneous with the magnetic field deflection. There is also a simultaneous increase in the transverse component (being the square root of the sum of squared $\theta$ and $\phi$ components) leading to a higher increase in the velocity than the radial component during the magnetic deflection. The velocity enhancement is also consistent with the SB statistics \citep[see Fig. 4 of][]{Larosa21}.
The magnetic deflection thus clearly displays an Alfvénic behaviour, with a link between the magnetic field and velocity field variations. These characteristics are consistent with a switchback signature. 

\begin{figure}
  \centering
   \resizebox{0.9\hsize}{!}{\includegraphics{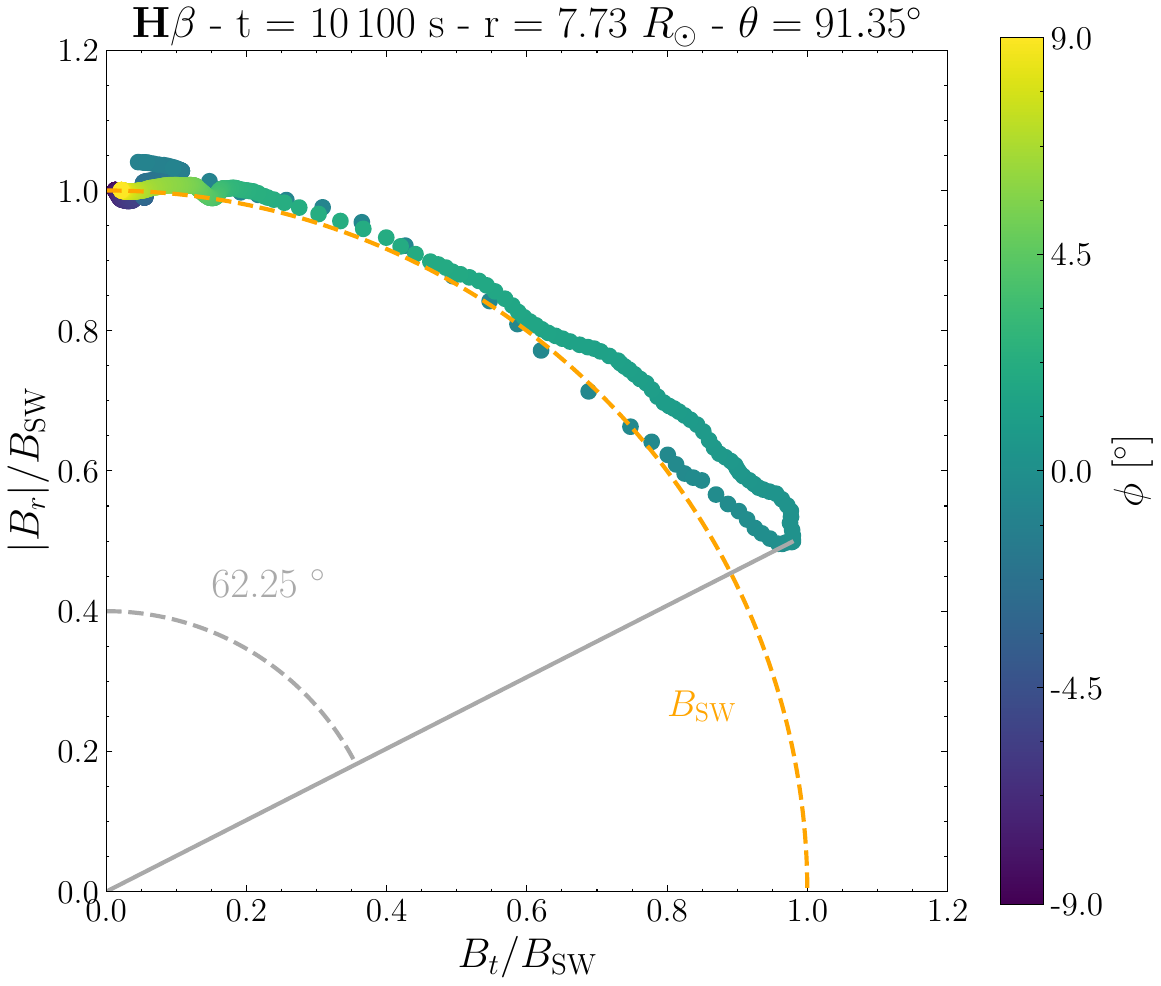}
   }
  \caption{Magnetic field evolution in the $B_t-B_r$ plane of the synthetic measurements of \fig{mag_deflections}. This normalised hodogram displays the change of the magnetic field vector orientation, colour coded according to the varying values of the $\phi$ coordinate. The quarter of constant ambient solar wind magnetic field intensity, $|B_{\mathrm{SW}}|$, is plotted with an orange dashed curve. The vector of maximum magnetic deflection is represented in grey.}
  \label{fig:Br_Bt}
\end{figure}

\begin{figure*}
  \centering
 \resizebox{0.9\hsize}{!}{\includegraphics{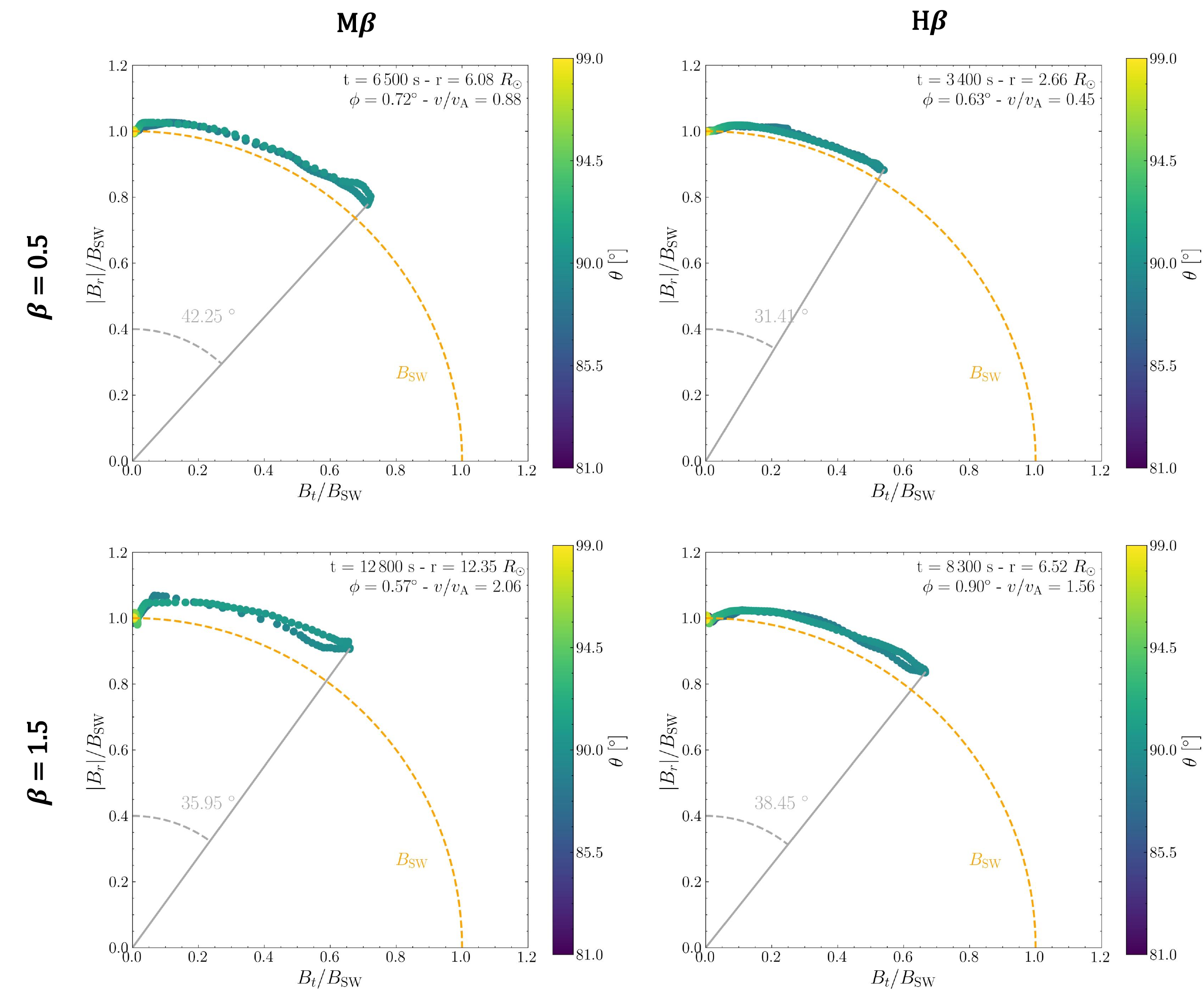}} 
  \caption{Characteristic hodograms of the magnetic field vector in the $B_t-B_r$ plane of M$\beta$ (left panels) and H$\beta$ (right panels) simulations at radial layers of two given values of the ambient plasma $\beta$: $\beta = 0.5$ (top panels) and $\beta = 1.5$ (bottom panels). The hodograms are built similarly to the one in \fig{Br_Bt}, with the difference that the dependence is on $\theta$ instead of $\phi$.}
  \label{fig:Br_Bt_beta}
\end{figure*}

Thus, the leading part of the jet, associated with an Alfvénic wave, exhibits signatures consistent with switchbacks. We estimate the maximum angle of magnetic deflection by plotting the magnetic field vector on a hodogram in the $Bt-Br$ plane (see \fig{Br_Bt}). The magnetic field vector $\vec{B}$, evolves on a sphere of nearly constant field intensity of the background solar wind, $|B_{\mathrm{SW}}|$. We define the deflection angle as the maximum deviation of the magnetic field vector, $\vec{B}$ from the radial direction, which represents the background magnetic field direction in our parametric simulations. The deflection angle for the synthetic in situ data of the lower panel of \fig{2Dcuts_evolution}, is approximately $62\degr$. This magnetic deflection, while not representing a full reversal, is thus substantial. Indeed, this deflection stands out from the background. Furthermore, it suggests the possibility of observing deflections without reversal — smaller deflections that exhibit a signature consistent with that of a switchback. This finding, along with other results presented in \fig{Br_Bt_beta}, indicates that smaller deflections may actually be predominant (see \citet{DudokdeWit20}).

Next, we briefly examine the magnetic deflections across different simulations and their evolution at different radial distances. \Fig{Br_Bt_beta} displays representative hodograms of magnetic deflections as a function of $\theta$ (i.e. at given $r$ and $\phi$), near the front of the torsional magnetic wave. The hodograms are obtained from both M$\beta$ and H$\beta$ simulations, where the torsional wave propagates through domains with specified plasma $\beta$. 
In the M$\beta$ case, the leading part of the jet, the untwisting magnetic wave crosses layers where the ambient plasma $\beta$ is $\sim 0.5$ at $t = 6\,500$ s. The hodogram taken at $r = 6.08 \, R_{\sun}$ and $\phi = 0.72\degr$ and displayed in the upper left panel of Fig. \ref{fig:Br_Bt_beta}, shows a deflection angle of about $42 \degr$. Later, the wavefront reaches the height where the background plasma $\beta$ $\sim 1.5$ at around $t = 12\,800$ s. The hodogram at $12.35 \,R_{\sun}$ and $\phi = 0.57 \degr$ (cf. Fig. \ref{fig:Br_Bt_beta} lower left panel) gives a deflection angle of $\sim 36 \degr$. This suggests that in M$\beta$, the magnetic deflection remains relatively constant, if not diminishing slightly. For the specific condition of this simulation there seems to be now growth of the deflection. 
In the H$\beta$ case, the leading part of the jet, the untwisting magnetic wave crosses layers where the ambient plasma $\beta$ is $\sim 0.5$ at $t = 3\,400$ s. The hodogram recorded at $r = 2.66 \, R_{\sun}$ and $\phi = 0.63\degr$ (upper right panel of Fig. \ref{fig:Br_Bt_beta}) indicates a deflection angle of about $31 \degr$. The wavefront subsequently reaches the plasma $\beta$ $\sim 1.5$ layer, around $t = 6\,500$ s. The hodogram obtained at $6.52 \,R_{\sun}$ and $\phi = 0.9\degr$ (cf. Fig. \ref{fig:Br_Bt_beta} lower right panel) shows a deflection angle of $\sim 38 \degr$. 
\Fig{Br_Bt} shows that when the wavefront is at $r = 7.73 \, R_{\sun}$, corresponding to an ambient plasma $\beta$ of $\sim 1.7$, the deflection angle is even higher ($\sim 62\degr$). Thus, in H$\beta$, it seems that the deflection increases as the untwisting wave propagates upwards and encounters layers with higher plasma $\beta$. This behaviour is not observed in M$\beta$.

To conclude, the analysis of magnetic deflection during the jet propagation in three parametric simulations reveals significant insights into its evolution over time and under varying plasma $\beta$ conditions. The leading part of the jet, associated with an Alfvénic torsional wave packet, demonstrates properties consistent with switchbacks, including a notable deflection in the radial magnetic field and a velocity enhancement. The observed deflection angles vary depending on the plasma $\beta$ value. For M$\beta$, the observed deflection angle seems to decrease with increasing ambient plasma $\beta$ value, whereas for H$\beta$, the observed deflection angle increases with increasing ambient plasma $\beta$ value.
These findings suggest that the dynamics of magnetic deflection are intricately linked to the plasma environment. A more comprehensive study of the deflection angle evolution in relation to atmospheric profiles, extending beyond the scope of this current study, shall be conducted in a future work. Future studies should also account for the role of the Alfvén speed and Mach number.

\section{Conclusions} \label{sec:Conclusion}

\subsection{Summary} \label{ssec:summary}

The present study investigates the dynamics of coronal jet propagation into the solar wind using adaptive 3D MHD simulations. 
Previous research by \citet{Pariat16} examined the influence of the plasma $\beta$ on jet propagation, focusing on a domain constrained to a small fraction of the solar radius above the surface and assuming a uniform atmosphere, without gravity and solar wind effects. In contrast, \citetalias{Karpen17} and \citet{Roberts18} investigated the propagation of a single jet within a stratified atmosphere over an extended domain, although in these studies, the jet remained within the sub-Alfvénic region, within a plasma $\beta \ll 1$ region. Building on these simulations, we conducted a parametric analysis to explore how different coronal radial plasma $\beta$ profiles can affect the jet propagation in a stratified atmosphere. The jet was generated impulsively and self-consistently through shearing motions at the photospheric boundary, in reference to the model of \citet{Pariat09a}. Additionally, the atmosphere was initialised using a uni-thermal Parker solar wind solution (\sect{sw}).

We confirm the results of \citet{Pariat16} that a self-consistent jet can indeed be generated and propagate in different low corona plasma $\beta$ conditions (\sect{param_sim}). After imposing a relaxation period and some photospheric shearing flows, we observed distinct stages such as the magnetic helicity and energy accumulation in the close parasitic polarity, the impulsive release of energy permitted by magnetic reconnection, and the sequential launch of untwisting magnetic torsional waves along newly formed field lines - across all simulations (\sect{Steps}). This suggests a robust and self-consistent generation model for coronal jets. Unlike prior works \citep{Pariat16, Karpen17, Roberts18}, which do not consider an Alfvénic surface, we explored the propagation of coronal jets from the sub-Alfvénic wind to the super-Alfvénic wind regimes in two simulations (L$\beta$ simulation serving as a baseline). 

As demonstrated by various numerical models \citep[\citetalias{Karpen17}]{Lionello16, Szente17}, our investigation confirms that the output of coronal jets can propagate easily over tens of solar radii away from the solar surface. Our parametric study shows that jet propagation can occur over a wide range of radial profiles of plasma $\beta$, with the Alfvén surface located at different distances, hence encompassing a wide range of conditions existing in the solar atmosphere \citep{West23}. Notable structures, such as the leading Alfvénic torsional wave and the trailing dense jet, were consistently observed, aligning with previous studies \citep{Pariat16, Uritsky17, Roberts18}. The leading structure corresponds to an Alfvénic torsional wave, characterised by a sudden increase in the velocity magnitude and a phase speed close to the local Alfvén speed, with a slight variation in mass density, indicative of its Alfvénic nature and strong rotational speeds. Conversely, patches of lower velocity, but higher mass density variations denote a dense bulk flow of plasma in the trailing dense-jet region (\sect{jet_propagation}).

While all structures are present in the parametric simulations, we observed differences in their propagation properties depending on the stratification profiles. Specifically, the phase velocity associated with the leading structure decreases from the low to the high $\beta$ simulations, reflecting differences in the radial ambient Alfvén velocity profiles. Additionally, the radial propagation speed ratio between the torsional magnetic wave and the dense bulk flow diminishes from the low to the high $\beta$ simulations. As a result, the two structures follow each other more closely with a higher $\beta$ profile (\sect{jet_propagation}).

To address whether the magnetic untwisting wave induced by a jet-like event can produce signatures akin to SBs or full reversal SBs, our three parametric simulations consistently show that the magnetic untwisting wave persists along magnetic field lines in the leading part of the jet (\sect{jet_dynamics}). Moreover, this jet-induced untwisting torsional magnetic wave exhibits characteristics typical of switchbacks:
\begin{itemize}
    \item minor variations in the magnetic field norm (less than 15\%);
    \item a deflection in the radial magnetic field, a decrease in the radial component, and an increase in the transverse component; and
    \item a simultaneous increase in radial velocity.
\end{itemize}
The angle of magnetic deflection ranges from $17\degr$ to $67\degr$ away from the ambient magnetic field, which is radial (\sect{mag_deflections}).

However, the three simulations show that the 'U-loops' observed close to the solar surface do not persist higher up in the corona (\sect{mag_deflections}). Moreover, in none of our parametric simulations were full reversal SBs' signatures observed, despite the presence of U-shaped loops at the jet onset (\sect{jet_propagation}).

Furthermore, the evolution of the magnetic deflection varies between simulations: in the M$\beta$ simulation, the deflection angle decreases as plasma $\beta$ radially increases from 0.5 to 1.5. Conversely, in the H$\beta$ simulation, the deflection angle increases over time, with propagation distance (radius) and ambient solar wind plasma $\beta$ (\sect{mag_deflections}).
Our parametric study thus deterministically demonstrates the ability of the jet-associated untwisting torsional magnetic wave to exhibit in situ characteristics measurements akin to switchbacks under diverse solar wind conditions.

\subsection{Discussion}\label{sec:Disc}

While our simulations provide valuable insights into the dynamics of the jets and the jet-associated magnetic untwisting waves  that induce typical switchbacks signatures, several limitations should be acknowledged.

The first point to consider pertains to the simulation's setup. In our study, the magnetic sources generating the jets have a spatial width of about $50$ Mm, corresponding to large-scale coronal jets. However, preliminary observational statistical studies indicate that such large coronal jets are likely too infrequent to fully explain the occurrence of SBs \citep{HuangN23}. 
Smaller-scale jet-like events, such as jetlets or spicules \citep{Kumar23, Raouafi23, Lee24}, are more likely candidates. To investigate this further, we would need to reduce the size of the jets to determine if the spatial scaling significantly influences the propagation properties and the ability of the jet-like outputs to reach the super-Alfvénic wind. This would require higher spatial resolution simulations, which is challenging due to the already substantial data volume of our runs. However, it is important to note the robust ability of the \citet{Pariat09a} model to generate jets under various conditions, as demonstrated by numerous studies (see \sect{Introduction}). This model's capability in a broad range of plasma $\beta$ environments  suggests that smaller-scale jets are likely to behave similarly to larger ones, indicating that changing the size may not affect the underlying mechanisms of jet formation.

Another limitation of the present simulations is the reliance on the uniform Parker model. The approach does not include an explicit energy equation, inhibiting modelling of certain physical processes. Our model does not account for other jet-driving mechanisms, such as evaporation flows \citep{Shimojo01}. Unlike the simulations by \citet{Lionello16} and \citet{Szente17}, which include more self-consistent solar wind generation mechanisms, our model employs only a simplified background solar wind. The advantage of this approach is the high level of tunability, which allows multiple parametric simulations to be carried out. Future development should incorporate more comprehensive solar wind model, such as polytropic ones \citep{keppens99}.

Additionally, it is important to consider the limitations of the ideal MHD paradigm used in our simulations. The fluid approximations employed are not ideally suited for the plasma conditions in the upper part of the domain. In the upper corona and inner heliosphere, the solar wind becomes collisionless, with particle mean free paths of the order of $1$ AU \citep{Verscharen19}. As a result, kinetic effects, which are known to be significant in the early solar wind, are completely neglected in our simulations.

The MHD treatment is considered both consistent and reliable for modelling the dynamics of large-scale eruptive events in the low solar corona. Our simulations highlight critical aspects of the solar origin scenario for switchbacks. Given that switchbacks are characterised as Alfvénic magnetic deflections with no characteristic deflection angle \citep{DudokdeWit20}, a primary outcome of our study is that the untwisting magnetic torsional wave packet induced through the generation of jets exhibits an in situ signature consistent with switchbacks (see \sect{mag_deflections}). Our simulations deterministically demonstrate a possible link between jet-like events and switchbacks, with magnetic deflections exceeding $62\degr$ observed in the super-Alfvénic wind. 

However, our analysis remains preliminary. In \sect{mag_deflections} we observed distinct radial variations in the properties of the magnetic deflection across two simulations with different radial profiles for $\beta$. Further investigation is needed to determine the driving forces behind the propagating magnetic deflection and whether the steepening of the torsional Alfvénic wave is a systematic phenomenon, as suggested by Alfvén wave growth through expansion mechanisms \citep{Squire20, Squire22, Shoda21, Johnston22}, or if it varies, as indicating by M$\beta$ simulation. Additionally, a more detailed examination of switchback characteristics, such as the nature of their boundaries as rotational or  tangential, is warranted \citep{Larosa21, Akhavan21, Akhavan22, Bizien23}. 

Focusing on the formation of full-reversal switchbacks, and in agreement with and complementing the simulations by \citet{Wyper22}, our analysis rules out the direct-injection scenario, where magnetic inversion or 'U-shaped' loops are transported directly from the low atmosphere into the super-Alfvénic solar wind. In none of our simulations, which cover a relatively broad range of stratified atmospheric profiles, did we observe the advection of 'U-shaped' loops — defined as magnetic deflections greater than $90\degr$ — as depicted in several conceptual models of switchback formation \citep[e.g.][]{Fisk20, Akhavan22}. Although, 'U-loops' are present at the onset of jets and during the very early stages of untwisting magnetic waves, the strong Lorentz forces prevalent in the low-$\beta$ corona immediately straighten these 'U-loops', preventing their propagation into the upper corona. Thus, our model suggests that full-reversal switchbacks are unlikely to be observed, hence infrequent, in the low-$\beta$ / sub-Alfvénic solar wind, which is consistent with initial observational analyses of the sub-Alfvénic region reported by PSP \citep{Bandyopadhyay22, Jagarlamudi23, Cheng24}.

Our simulations suggest a two-step scenario for the formation of full-reversal SBs. Initially, jet-like events generate propagating torsional magnetic waves within the low-beta corona. These waves induce Alfvénic magnetic deflections, which can directly account for the majority of SBs observed. Some of these deflections may then serve as perturbation seeds for secondary in situ processes, leading to the steepening and eventual reversal of the magnetic field, thereby forming full-reversal SBs. Potential secondary mechanisms include Alfvénic expansion growth \citep{Squire20, Squire22, Shoda21, Johnston22} or velocity shears \citep{Toth23}. This scenario integrates a solar origin for the initial seeds, suggesting that the statistical properties of SBs are influenced by the distribution and frequency of solar-like jets, while in situ mechanisms are responsible for generating full-inversion SBs and determining some of their specific properties.

To better understand the properties, dynamics, and radial evolution of these seed Alfvénic waves, we need to enhance the production of synthetic in situ data from our simulations. In this paper, we assumed that the torsional wave is intersected by an infinitely fast spacecraft. To improve the realism and accuracy of comparisons between synthetic and observed data, we must implement more realistic spacecraft trajectories, similar to those described by \citet{Lynch21}. This likely necessitates rerunning the simulations with a higher graphics output cadence. Such improvements will enable a more accurate validation and refinement of our simulation models against observations from PSP and Solar Orbiter. 

Overall, a significant gap in understanding the propagating dynamics of jets arises from observational limitations, including the sampling of the middle corona, spatial resolution, cadence, and access to lower regions \citep{West23}. New coronagraphic observations, such as those enabled by METIS/Solar Orbiter \citep{Antonucci20}, can provide valuable new insights. The S-shaped structures observed in METIS data \citep{Telloni22} may correspond to signatures of large-scale propagating jets. The study of propagation is particularly important for statistical analyses attempting to link low coronal observations with SB observations. To further investigate these phenomena, observational campaigns could be designed to detect similar signatures near the solar surface with METIS and track their propagation using instruments on PSP and Solar Orbiter. Additionally, the upcoming PROBA-3 \citep{Zhukov16} and Polarimeter to UNify the Corona and the Heliosphere (PUNCH) missions \citep{Deforest22, Cranmer23}, with their next-generation coronagraphic capabilities, are expected to offer groundbreaking insights into the propagation of intermediate and small-scale jet events towards the super-Alfvénic wind.

\begin{acknowledgements}

We thank the anonymous referee for her/his thorough review of the manuscript.
JT acknowledges funding from the Initiative Physique des Infinis of Sorbonne Université. This work was supported by the French Programme National PNST of CNRS/INSU, co-funded by CNES and CEA, and by financial support from the French national space agency (CNES) through the APR program. Essential tests and preparatory simulation runs were made possible thanks to the granted access to the HPC resources of MesoPSL, financed by the Region Île-de-France and the project Equip@Meso (reference ANR-10-EQPX-29-01) of the programme Investissements d’Avenir, supervised by the Agence Nationale de la Recherche. This work was also granted access to the HPC resources of IDRIS and CINES under the allocations A0130406331 and A0160406331 made by GENCI, which enabled the production simulation runs.  CF acknowledges funding from the CEFIPRA Research Project No. 6904-2. PW was supported by an STFC (UK) consortium grant ST/W00108X/1 and a Leverhulme Trust Research Project grant. The authors thank Rick DeVore, Benjamin Lynch and Judy Karpen for their physical insights and useful discussions.

\end{acknowledgements}

%
%

\end{document}